\documentclass[structabstract,a4paper]{aa}  
\usepackage{graphicx}
\usepackage{txfonts}
\usepackage{rotating}
\begin{document}
\title{Averaging the AGN X-ray spectra from deep Chandra fields}

\subtitle{}

\author{S. Falocco 
  \inst{1}
  \and
  F. J. Carrera
  \inst{1}
  \and
  A. Corral
  \inst{5}
  \and
  E. Laird
  \inst{2}
  \and
  K. Nandra
  \inst{3,2}
  \and
  X. Barcons
  \inst{1}
  \and
  M. J.Page
  \inst{4}
  \and
  J. Digby-North
  \inst{2}
}
\institute{Instituto de F\'isica de Cantabria (CSIC-UC)
   39005 Santander, Spain \\
  \email{faloccco@ifca.unican.es}
  \and
  Astropysics Group, Imperial College London, Blackett Laboratory, Prince Consort Road, London, SW7 2AZ, UK 
  \and
  Max Planck Institute f\"ur Extraterrestriche Physik. 85748 Garching, Germany
  \and
  Mullard Space Science Laboratory, University College London, Holmbury St. Mary, Dorking , Surrey RH5 6NT, UK
  \and
  Osservatorio Astronomico di Brera INAF, Via Brera 28, CAP 20121, Milano, Italy
}

\date{Accepted on 10 November 2011  }


\abstract
{The X-ray spectra of Active Galactic Nuclei (AGN) carry the
  signatures of the emission from the central region, close to the
  Super Massive Black Hole (SMBH). For this reason, the study of deep
  X-ray spectra is a powerful instrument to investigate the origin of
  their emission. }
{The emission line most often observed in the X-ray spectra of AGN
      is Fe K$_{\mathrm{\alpha}}$.  It is known that it can be broadened
      and deformed by relativistic effects if emitted close enough to the
      central SMBH. In recent statistical studies of the X -ray
      spectra of AGN samples, it is found that a narrow Fe line is
      ubiquitous, while whether the broad features are as common is
      still uncertain. We present here the results of an investigation on the
      characteristics of the Fe line in the average X-ray spectra of AGN
      in deep Chandra fields.}
{The average spectrum of the AGN is computed using \emph{Chandra}
      spectra with more than 200 net counts from the AEGIS,
      Chandra Deep Field North (CDF-N) and Chandra Deep Field South
      (CDF-S) surveys. The sample spans a broader range of X-ray
        luminosities than other samples studied with stacking methods
        up to $z\sim3.5$.  We analyze the average spectra of
        this sample using our own averaging method, checking the results
        against extensive simulations. Subsamples defined in
      terms of column density of the local absorber, redshift and luminosity are also
      investigated.}
{We found a very significant Fe line with a narrow profile in all our samples and in almost all
    the subsamples that we constructed. The equivalent width of the narrow line estimated in the average spectrum of the full sample is 74 eV. The broad
    component of the Fe line is significantly detected in the
    subsample of AGN with $L < 1.43
    \times 10^{44} $ erg s$^{-1}$ and $z < 0.76 $, with an equivalent width 108 eV. }
{ We concluded that the narrow
    Fe line is an ubiquitous feature of the X-ray
    spectra of AGN up to $\langle z \rangle \sim 3.5$.  The broad
    component of the Fe line is significantly detected in the X-ray spectra of the AGN with low luminosity and low reshift. }
\maketitle

\section{Introduction}
\begin{table*}[h]\label{tproperties200}
  \small
  \caption{Properties of the full sample and the subsamples (see text).}
\begin{tabular}{lrrrrrrlllll}
\hline
Sample &  $N$  &   $N_{2-12}$  & $N_{5-8}$   & $\langle z \rangle$ & $\langle L_{43} \rangle $  & $\langle N_{h,22} \rangle$ &   $\Gamma$  &   $\chi^2$/dof  & $\Gamma_{sim}$ & $\sigma_{sim}$ & EW$_{sim}$ \\  
 &  &  &    &  & $10^{43}  $ erg s$^{-1}$  & $10^{22}$ cm$^{-2}$  &    &  & &  eV & eV    \\  
 (1) & (2) & (3) & (4)   & (5) &  (6)  & (7) &  (8)  &     (9) & (10) & (11) & (12)\\
\hline
 
 \textbf{CDF-S}     &   33 &  21561  &  5223    & 1.26   & 5.20    &  3.80  &   - &  - & - & - &-  \\ 
   
 \textbf{CDF-N}     & 25  &  16134  &  3429   &   1.11    & 8.00 & 0.88  &   -   &   -  & - & - &-  \\ 

 \textbf{AEGIS}     & 65  &  32971  & 6873 &   1.11 & 20.31 &   1.04  &   -  &   -  & - & - &-   \\ 
\hline
 \textbf{Total}   & 123 &  70667 &  15526 & 1.15     & 13.82   &    1.75  & $1.24\pm 0.05$   & 9.77/13  & $1.897\pm 0.001$ & $117\pm 1$ &  182 \\
 \hline
\textbf{$\mathrm{log{(N_{\rm H})}}>21.5$}    & 54 & 27888  & 7304  &   1.15   &	14.11	 & 3.93   & $0.48 \pm 0.07 $   & 8.94/13 & $1.898\pm 0.001$ & $129\pm 1$ & 190  \\ 
\textbf{$\mathrm{log{(N_{\rm H})}}<21.5$}   &  69   & 42778  & 8222  &   1.14   &  	13.55	  & 0.05 & $1.78\pm 0.06$   & 14.96/13  & $1.896\pm 0.001$ & $116\pm 1$ & 181   \\ 
\hline
\textbf{ $\mathrm L_{43} < 8 $}   & 74 & 34620  &  7630 &  0.85    & 2.89  & 1.64  & $1.21 \pm 0.06$   & 10.42/13  &  $1.893 \pm 0.001$ & $111\pm 1$ &  183  \\ 

\textbf{$\mathrm L_{43} > 8 $}       & 49 & 36047  & 7896  &  1.60    & 	30.20		 & 1.91   & $1.31 \pm 0.07$   & 14.70/13 &  $1.878\pm 0.001$ & $128\pm 1$ & 174  \\ 
\hline
\textbf{ $\mathrm z <  1.005$}            & 63 & 34843  & 7507  &  0.70    & 	5.43		 &      1.28     & $1.23 \pm 0.06$   &   10.57/13 & $1.896\pm 0.001$ & $105\pm 1$ &  186 \\ 

\textbf{ $\mathrm z >  1.005$}         & 60 & 35824  &  8019 &  1.62    & 	22.50 & 2.24   & $1.24 \pm 0.07$   &  5.66/13 & $1.886\pm 0.001$ & $138\pm 1$ & 180  \\ 
\hline
\textbf{ $\mathrm z <  0.76$}     & 39 & 23713  & 5285  &  0.57    & 	
3.23	 & 1.17   & $1.31 \pm 0.07$   & 11.68/13  & -$_{}^{}$ & -$_{}^{}$ &-   \\ 

\textbf{$\mathrm L_{43} < 14.3   $, $\mathrm z < 0.76 $}       & 36 & 19007    &  4446 &  0.57   & 		
2.00	 &  1.25    & $1.29 \pm 0.08$   & 10.99/13   & -$_{}^{}$ & -$_{}^{}$ &-    \\ 

\textbf{$\mathrm L_{43} < 14.3  $, $\mathrm z > 0.76 $}     & 49 & 24286  & 4902  &  1.14    & 	5.58	 &  1.90  & $1.11 \pm 0.08 $   & 7.27/13  & -$_{}^{}$ & -$_{}^{}$ &-  \\ 

\textbf{$\mathrm L_{43} > 14.3 $, $\mathrm z > 0.76 $}             &  35   & 22669  & 5339  & 1.80     & 37.05		 &  2.19  & $1.34 \pm 0.09 $   &  11.76/13 & -$_{}^{}$ & -$_{}^{}$ &-  \\ 

\hline

\hline
\end{tabular}

\tablefoot{Columns: (1): (Sub)Sample; (2): Number of sources; (3): number of net counts in 2-12 keV rest-frame; (4): number of net counts in 5-8 keV rest frame; (5): average redshift; (6): average  rest-frame 2-10~keV luminosity in units of $10^{43}$ erg s$^{-1}$, corrected for Galactic and intrinsic absorption$^*$; (7): average intrinsic column density in $10^{22}$ $\mathrm{cm^{-2}}$ $^*$; (8): gamma from the fit of the average observed spectrum with a powerlaw between 2 and 5 keV (rest-frame); (9): $\chi^{2}$/dof of the fit of the average observed spectrum with a powerlaw between 2 and 5 keV (rest-frame); (10): slope of the powerlaw  $^{**}$ of the average spectrum obtained with the simulations of the Fe line (see text); (11): width obtained in those simulations; (12): EW of the line obtained from those simulations.
$^*$ Obtained averaging the values obtained from the fit to the individual spectra (see text); $^{**}$ obtained fitting it with a powerlaw plus a Gaussian Fe line}
\end{table*}
The X-ray spectra of AGN are emitted from the innermost regions of the
central engine, close to the central SMBH. Accordingly to the
currently accepted model described in detail by \cite{shakura}, AGN
are powered by accretion to the SMBH, with large amounts of potential
energy released in the accretion disc. Assuming an optically thick,
geometrically thin accretion disc, the optical-UV primary emission of
the AGN is explained as the result of the thermal emission from the
accretion disc.

It is well known that the emission in hard X-rays is dominated by a
powerlaw with $\mathrm \Gamma$ $\sim 2$, and the
most widely accepted interpretation for it invokes a hot plasma
surrounding the accretion disc that makes inverse Compton scattering
of the thermal optical-UV photons from the accretion disc
(\cite{haardt}).

Part of the primary emission is reflected by the accretion disc
producing Compton reflection (especially significant above 10 keV)
and several fluorescence lines, the most important one being the Fe
$\mathrm K_{\alpha}$ line at 6.4 keV for neutral Fe (and 6.7-6.9 keV for Fe in the most highly ionized states), as described in
\cite{matt91} and \cite{george91}. Reflection can also occur at larger
scales, in the torus. The line profile gives valuable information
about the emitting processes and the regions where they occur.

The study of the trend of the narrow Fe line equivalent width (EW) with
continuum luminosity plays a fundamental role in the investigation of
the link between the continuum emitting region and the reflecting
region. A decrease of Fe line intensity with increasing
continuum luminosity was found in Ginga observations of AGN (
\cite{iwasawa}).  The trend (the so called "Iwasawa Taniguchi effect",
hereafter ``IT effect'') was widely confirmed for both broad and narrow Fe lines by observations of \emph{ASCA}, by
\cite{nandra97}, and more recently for the narrow Fe lines, for example in \cite{bianchi2007}
and \cite{page}, using \emph{XMM-Newton} data.  Although the IT effect is widely proved, the reason
for the anti-correlation between the line and the continuum intensity
is still under investigation: \cite{nandra97} attributed the IT effect to the
ionization that can be important in high luminosity.

As the torus and the disc can both contribute in different measure to
the observed Fe line, the resulting profile will be complex, and in particular it
will depend on the inclination angle of the disc with respect to the
line of sight. The estimated intrinsic column density in X-rays ($\rm
N_{\rm H}$) is, in this sense, connected to the line properties: a
increase of the Fe line intensity with increasing $\rm N_{\rm H}$
was found in a sample of AGN observed by \emph{Ginga} and \emph{Asca}
(\cite{gilli}). In that work, the Fe line in the spectra of the
absorbed AGN (type 2 AGN) was more intense than the line found in the
unabsorbed ones (type 1 AGN). This was predicted by \cite{ghisellini},
who performed a theoretical analysis of the Fe line in a
configuration with torus and accretion disc on the same plane. In this
picture, the line intensity produced by reflection and transmission in
the torus increases with the $\rm N_{\rm H}$, while the continuum
emission is depleted, leading to an increase of the line equivalent
width for the highest $\rm N_{\rm H}$.

It should be noted that the lines produced by reflection in the torus
should have a narrow and approximately Gaussian profile, while
the lines produced in the accretion disc should be broadened and
deformed by the effect of the newtonian movement of the gas in the
accretion disc around the SMBH. It is also possible to detect in the lines the
relativistic features, such as a red wing due to gravitational redshift, among other features, explained in detail in
\cite{fabian2000}. These features come from the innermost regions of
the accretion disc, as the relativistic effects are due to the strong
gravitational field close to the central SMBH. 
From the study of these
features it is possible to obtain the inner radius of the accretion
disc, which is expected to be 6 gravitational 
radii ($R_{g}$) for a non-rotating BH, called a Schwarzschild BH, down to 1.23 $R_{g}$ 
for a maximally rotating BH, called a Kerr BH (\cite{diskline2}).

Early results from ASCA showed broad and deformed lines:
\cite{nandra97} performed a spectral analysis of a sample of local
Seyferts, finding broad components in about 65\% of their sample.
Unambiguous broad lines were found also in more recent well exposed XMM-Newton
observations of AGN (e.g. in \cite{nandra2006}, \cite{braito2007} and
\cite{fabian2002}).


It has been proved by \cite{Guainazzi} that X-ray spectra with
  good statistical quality are necessary to detect any broad Fe
line component, which would otherwise be hidden under the noise.
In this context statistical methods, like summing (stacking) or
  averaging the spectra, have been recently introduced in X-ray
astronomy to allow the study of large AGN samples including the
low quality spectra that otherwise could not be well analyzed
individually.

The first of such works is that of
\cite{alina}, who performed a stacking analysis on the
\emph{XMM-Newton} observation of the Lockmann Hole, finding a large line EW
 and broad line profile. 

\cite{brusa2005} stacked the spectra of CDF-N (2 Ms exposure
 time) and CDF-S (1 Ms exposure time) surveys by the
\emph{Chandra} satellite. They computed the stacked spectra in bins of
redshift aiming at characterizing the Fe line emission of the
sources in the X-ray background up to $ z\sim 4 $. They found an
intense and apparently broad 6.4 keV Fe line with an EW consistent
with the results of \cite{alina}. To explain the red component, they argued that it can strongly depend on the modeling of the underlying
continuum and that a spurious red wing might be produced by the
contribution of absorbed spectra at different redshifts.

In recent studies of X-ray spectra using this approach, finding
broad lines with relativistic profiles has been proved to be not very common, for example in
\cite{Guainazzi} in the analysis of AGN
spectra observed with \emph{XMM-Newton}. They found that 25\% of their sample had relativistic
lines; this percentage is 50\% for higher signal-to-noise ratio,
selecting only spectra with large numbers of counts. Moreover, they
found the strongest relativistic profiles in low
luminosity objects. 

\cite{corral2008} averaged the \emph{XMM-Newton} spectra of type 1 AGN, using the AXIS (\cite{mateos2005}) and XWAS (\cite{mateos2008}) samples, up to redshift $\sim$ 3.5. They found a
narrow Fe line, while no clear evidence of a broad line was found.
The reason for the discrepancies between this work and that of \cite{alina} can be
explained considering the differences of the samples, as the
\cite{corral2008} sample has higher luminosities (and therefore lower predicted EWs, following the IT effect) and less spectral counts,
including more noisy sources. Another reason for the discrepancy can be
found in the method, since \cite{corral2008} estimate the
  continuum shape using simulations, while in \cite{alina} it was not
constructed and subtracted in the same way and this could introduce
some uncertainties. Moreover, \cite{alina} binned the spectra before
stacking them, and \cite{Yaqoob2006} showed that this procedure can introduce
features like a broad red tail in an emission line.

A stacking analysis of a deep and complete sample of 507 AGN with
$z<4.5$ defined from the 2XMM catalogue was performed by
\cite{chaudhary2010}. They were able to characterize the properties of the
stacked spectrum of the AGN, such as the Fe line shape and the
dependence of its intensity with X-ray continuum luminosity and the
redshift. They found a clear evidence of a narrow neutral Fe line,
and they confirmed the IT effect in AGN over a broad range of
redshift.

Recently, a stacking analysis on {\emph XMM-Newton} X-ray spectra of the COSMOS
sample was performed (\cite{iwasawa2011}): they found an excess
on the high-energy side of the Fe line, interpreted as the convolution of narrow
lines from ionized Fe.

In summary, according to recent results, while the narrow Fe lines are
commonly detected in the spectra, clear evidence for relativistic
lines is rare. The paucity of relativistic lines is a problem, as they
are expected in the accretion disc scenario. There are a number of
possible solutions for this problem. The first one invokes the
presence of ionized discs (\cite{iwasawa2011}, \cite{matt96}); the
second solution explains it with the presence of fastly spinning BH
(\cite{iwasawa96} and \cite{fabian2002}): in this case the lines would
be very broad and sometimes difficult to separate from the continuum.
The last explanation for the lack of broad lines postulates truncated
discs, as found by \cite{matt2005}.

We present in this paper a stacking analysis of a comprehensive sample
of absorbed and unabsorbed AGN in the deepest \emph{Chandra} surveys:
AEGIS and Chandra Deep Fields. We followed the method presented by
\cite{corral2008}, and we accurately tested and adapted it for
\emph{Chandra} spectra, in order to carefully check our results.
The \emph{Chandra} sample in this work allows us to explore 
the properties of AGN with the best statistics ever reached at high redshift, as it is deeper than the \emph{XMM-Newton} samples that have previously been analyzed with stacking techniques, such as
  \cite{corral2008} and \cite{chaudhary2010} (see Sect. 2.2). In
  comparison with the previous stacking analysis of the Chandra
  samples in \cite{brusa2005}, we include the sources from the AEGIS-X
  survey, improving the statistics of high luminosity AGN, and we use
  the more recent (2 Ms) observation of the CDFS (Sect. 2.1). 

The paper is organized as follows: the properties of the sample are
described in Sect. 2; the method in Sect. 3; discussion of the results
is in Sect. 4; we summarize the conclusions in Sect.
\ref{conclusions}.

Across this paper, we adopt the cosmological parameters: $H_{\ 0}$=70
km s$ ^{-1}$ Mpc$^{-1}$; $ \Omega_{\lambda}$=0.7
(\cite{komatsu}).

In this paper, all the counts refer to the net number of counts
between 2 and 12 keV rest-frame; all the luminosities are calculated
between 2 and 10 keV rest-frame and are corrected for Galactic and
intrinsic absorption using the fits to the individual spectra (see
below).  We have used \texttt{XSPEC v. 12.5} (\cite{xspec}) for
spectral analysis.


\section{X-ray samples}
The X-ray results presented in this paper are obtained from the \emph{Chandra} observations of the AEGIS, CDF-N and CDF-S surveys. 

\subsection{Parent Chandra surveys}

The AEGIS-X survey is a \emph{Chandra} survey of the Extended Groth
Strip (EGS) region, designed primarily for studying the co-evolution
of black holes and their host galaxies. The data we used are the
  result of 8 contiguous Advanced CCD Imaging Spectrometer (ACIS-I)
pointings, each with exposure 200 ks, totalling 1.6 Ms. The survey
covers a total area of approximately 2302 arcmin$\rm ^2$ a in a strip
of length 2 degrees.  The total number of identified sources with
spectroscopic redshifts is 409.  The data reduction and point source
detection algorithms used to analyze these data are described in
\cite{laird2009}.

The Chandra Deep Field-North (CDF-N) survey is one of the deepest
  (2Ms with ACIS-I) 0.5-8.0 keV surveys ever made: nearly 600 X-ray
  sources are detected over 448 arcmin$\rm ^2$. The total number of
sources identified with spectroscopic redshifts is 307. Details of
data reduction and the point source catalog are described in the paper by \cite{alex}.

The Chandra Deep Field South (CDF-S) is the deepest and most sensitive
observation obtained with \emph{Chandra}. We used the 2 Ms survey,
covering an area of 436 arcmin$\rm ^2$. Several hundred X-ray point
sources were detected. The number of sources identified with
spectroscopic redshifts is 152.  A detailed description of the survey
can be found in the paper \cite{luo2008}.  It has been recently
extended to 4 Ms, as described in \cite{alex2011}, and \cite{Xue}.

\subsection{Sample definition and properties}\label{sample_intro}


Our main purpose is to study the average spectrum of AGN in the 2-12
keV energy band (rest-frame), and to determine the shape of the Fe line
at $\sim$6.4 keV. The X-ray spectra of our sample have few counts and do not
allow a detailed analysis of the individual spectra, except for a
handful of sources. For the detection of any relativistically
broadened component, as underlined by \cite{Guainazzi}, it is necessary to have good quality spectra, hence we need to improve the signal-to-noise ratio.

To increase the probability of detecting any broad component that
otherwise would be hidden by the noise, we included in our analysis only those sources with more than 200 counts. We discuss in Appendix A
the results corresponding to lower thresholds (50 and 100 counts).
We further selected only sources with spectroscopic redshifts.
  Finally, we excluded from our analysis also two sources with more than 10000 counts individually: CDFS$_{-}$056 (RA: 53.112, DEC: -27.685), CDFN$_{-}$141 (RA: 189.1, DEC: 62.383). These sources have such a strong signal in their spectra that they would largely dominate the average stracked spectrum.

\begin{figure}
  \centering
  \includegraphics[width=9cm]{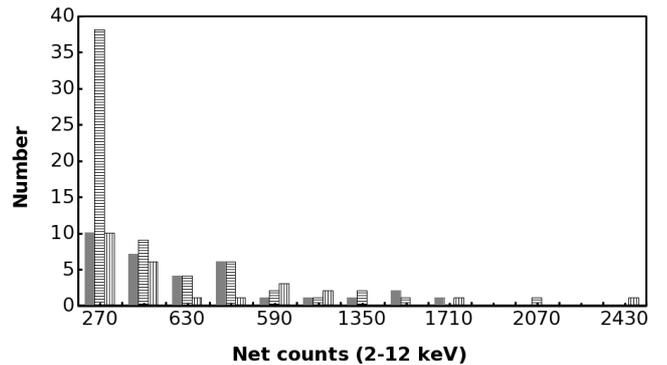}
  \caption{ Distribution of the net counts between 2 and 12 keV
    rest-frame. Filled histogram: CDF-S; histogram with horizontal stripes: AEGIS, histogram with vertical stripes: CDF-N.}
  \label{Fig:DistrCts}
\end{figure}

\begin{figure}
  \centering
  \includegraphics[width=9cm]{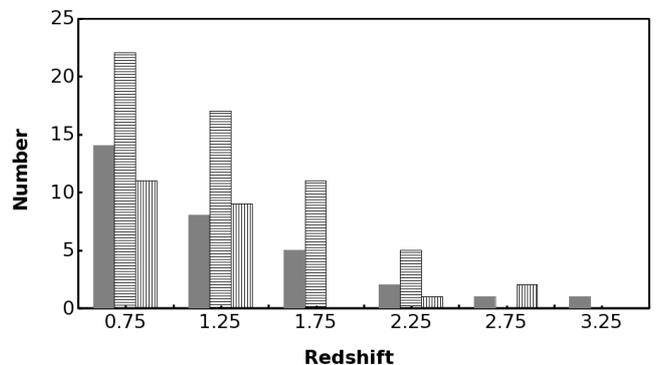}
  \caption{Redshift distribution. Filled histogram: CDF-S; histogram with horizontal stripes: AEGIS, histogram with vertical stripes: CDF-N.}
  \label{Fig:DistrZeta}
\end{figure}

Our full sample contains 123 sources with spectroscopic
redshifts and more than 200 counts each (but less than 10000 cts), with
70667 counts in total (see Table \ref{tproperties200}).  
The distribution of the net counts per source for this full sample is shown in Fig.\ref{Fig:DistrCts}.  The redshift distribution of the sources is
shown in Fig. \ref{Fig:DistrZeta}, where we can see that our sample
spans a broad range of redshifts from $z=0$ up to $z=3.5$.

 Source and background spectra, and their ancillary and response
  matrices were extracted for each source (\cite{jonathan}), using the
  tool \texttt{ACIS-EXTRACT}, version 2008-03-04, \cite{patrik}.
  Before spectral fitting, all the spectra were grouped with a minimum
  of 5 counts per bin, and we applied the modified Cash statistics discussed in the XSPEC pages \footnote{ http://heasarc.gsfc.nasa.gov/docs/xanadu/xspec/wstat.ps}. This grouping was adopted
  because we used the same method for all the sources with more than
  50 counts (see Appendix \ref{appendix_results}), for which the more common $\chi^2$
  statistics would not be adequate.

\begin{figure}
  \centering
  \includegraphics[width=9cm]{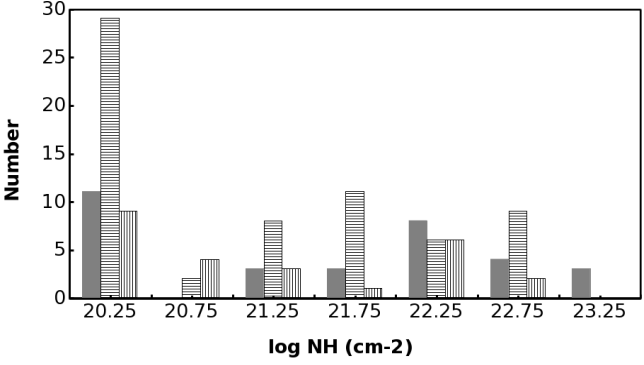}
  \caption{Distribution of intrinsic column density in units of $10^{22}$ cm$^{-2}$ . Filled histogram: CDF-S; histogram with horizontal stripes: AEGIS, histogram with vertical stripes: CDF-N. We grouped the spectra with $\rm log(N_{\rm H})<10^{20}$ with the spectra having $\rm log (N_{\rm H})=10^{20}$. }
  \label{Fig:DistrNh}
\end{figure}

It should be noted that in this step we did not aim at making a
detailed analysis of the individual spectra but rather only at
obtaining a set of parameters that we use  to unfold the spectra for all the instrumental effects (see Sect. \ref{method}).  The model is a
single power law modified by Galactic absorption (with fixed values at 1.4 $\times \rm 10^{20}\rm cm^{-2}$ for AEGIS, 0.772 $\times \rm 10^{20}cm^{-2}$ for CDF-S, 0.993 $\times \rm 10^{20} \rm cm^{-2}$ for CDF-N) 
plus intrinsic absorption at the redshift of each source.
We left the slope of the power law, its normalization and the
intrinsic column density as free parameters. The fit was performed in
the rest-frame energy range between 1 and 12 keV, minimizing the
contribution from any putative soft excess present in the sources. The resulting
distribution of intrinsic column densities is shown in
Fig.\ref{Fig:DistrNh}. Most of the highly absorbed sources belong to
CDF-S, as indicated by the higher average column density for that
sample in Table \ref{tproperties200}.


\subsection{Subsample definition}\label{description}


\begin{figure}
  \centering
  \includegraphics[width=9cm]{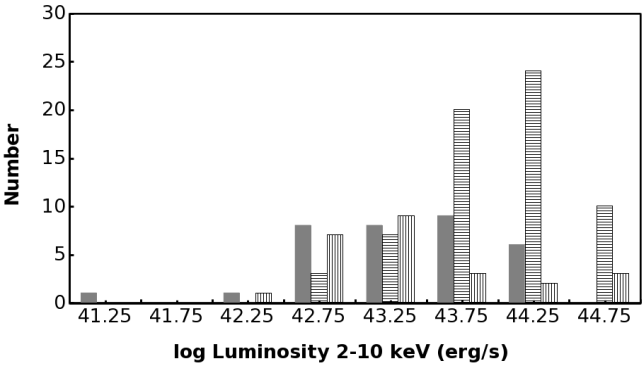}
  \caption{Distribution of rest-frame 2-10~keV luminosities in units of erg s$^{-1}$. Filled histogram: CDF-S; histogram with horizontal stripes: AEGIS, histogram with vertical stripes: CDF-N.}
  \label{Fig:DistL}
\end{figure}         

We show in Fig. \ref{Fig:DistL} the distribution of the X-ray
luminosities of the sources corrected for Galactic and intrinsic
absorption, calculated from the fit to the indvidual spectra. The
average luminosity of the AEGIS sources is one order of magnitude higher than the average luminosity of CDF-N and CDF-S sources (Table \ref{tproperties200}).

\begin{figure}
  \centering
  \includegraphics[width=6.5cm,angle=-90]{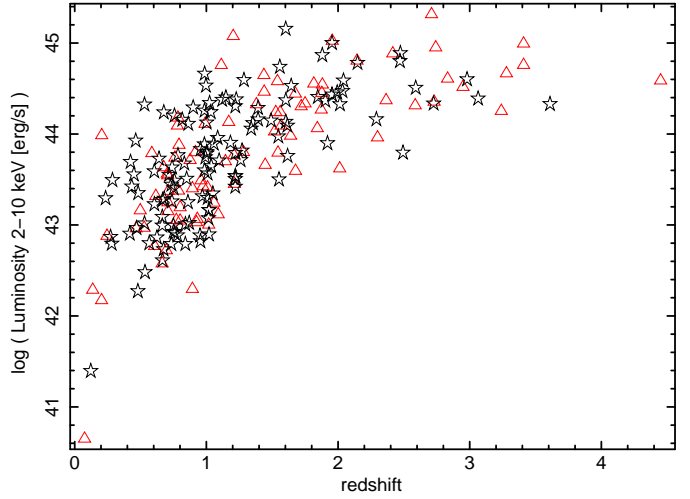}
  \caption{Comparison of our full sample (black stars) with the deep \emph{XMM-Newton} sample of \cite{alina} (LH, red triangles), in the luminosity-redshift plane.}
  \label{Fig:LzLH}
\end{figure}

\begin{figure}
  \centering
  \includegraphics[width=6.5cm,angle=-90]{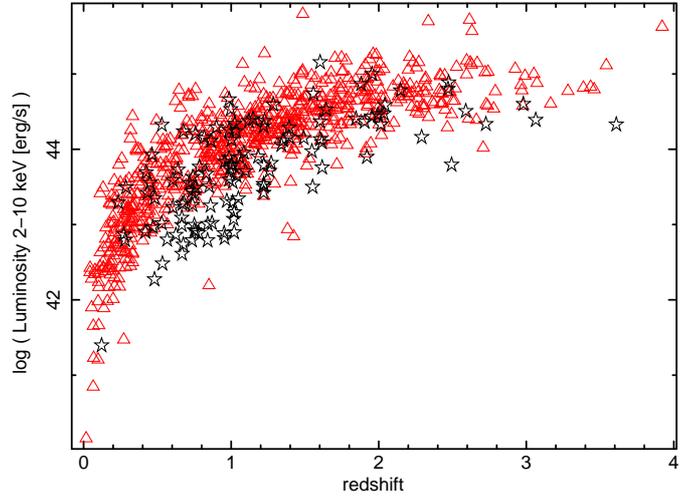}
  \caption{Comparison of our full sample (black stars) with the medium \emph{XMM-Newton} sample of \cite{corral2008} (XMS+XWAS, red triangles), in the luminosity-redfhit plane. For clarity, we omitted one XMS+XWAS source of XMS+XWAS at $z=2.34$, $L= 1.28 \times 10^{47}$ erg s$^{-1}$.}
  \label{Fig:LzXMS}
\end{figure}

We compared the distribution of our ful sample in the
luminosity-redshift plane with that of \cite{alina} in the Lockmann
Hole (Fig. \ref{Fig:LzLH}): we sample better that plane at $z<3$ for
all luminosities. Compared to the \cite{corral2008} sample (XMS-XWAS)
(Fig. \ref{Fig:LzXMS}) we cover about one order of magnitude lower
luminosities for $0.5<z<3$. Finally, with respect to \cite{brusa2005},
the addition of the AEGIS sample and the restriction to sources with
high numbers of counts, allows us to span a broader range of
luminosities at similar $z$. In summary, our sample combines a unique combination of deep coverage with high z sources, a broad span of luminosities and good statistics. 

%

%

\begin{figure}
  \centering
  \includegraphics[width=6.5cm,angle=-90]{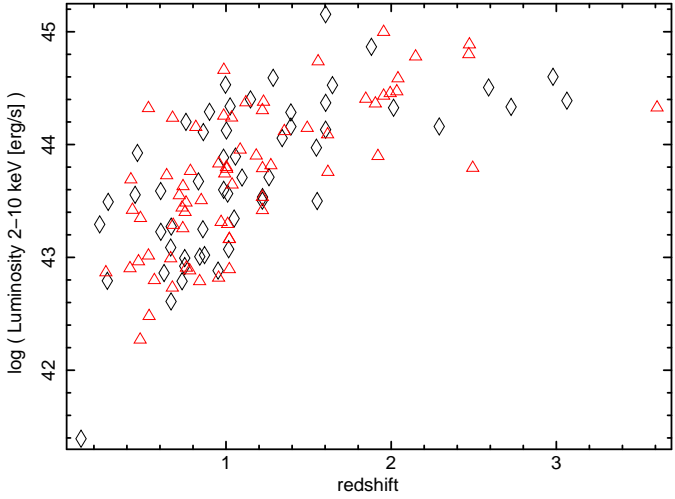}
  \caption{Distribution of the unabsorbed (red triangles) and absorbed (black diamonds) subsamples (see Sect. 2.3) in the luminosity-redshift plane.}
  \label{Fig:DistrLz_abs_unabs}
  \end{figure}

\begin{figure}
  \centering
  \includegraphics[width=6.5cm,angle=-90]{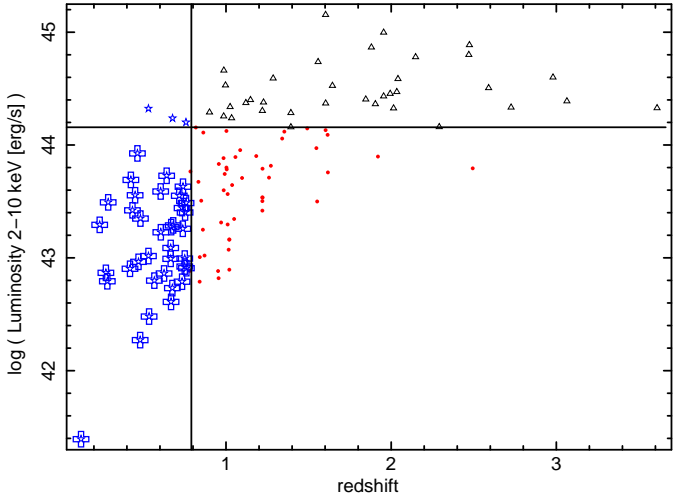}
  \caption{Distribution of the $L-z$ subsamples (see text) in the luminosity-redshift plane. The dividing point is $z=0.76$, $L=1.435\times10^{44}$~erg~s$^{-1}$ (marked by the horizontal and vertical solid lines). The three blue stars mark the sources that allow defining distinct $L-z$ subsamples (see text).}
  \label{Fig:DistrLz_bins}
\end{figure}

One of the main aims of our work is to understand the dependence
  of the spectral properties of the AGN on luminosity, redshift, and
  intrinsic absorption. For this reason, we performed the analysis of
  the stacked spectra not only for the full sample, but also for
  subsamples defined in terms of column density, luminosity, and redshift.
  Except for $N_{\rm H}$ (see below) the bins have been designed to
  have a similar numbers of total counts each. The characteristics of
  each subsample can be seen in Table~\ref{tproperties200}.

\begin{itemize}

\item{Intrinsic $N_{\rm H}$:} The
threshold column density was set at $\log(N_{\rm H}/{\rm
  cm}^{-2})=21.5$ as the absorption features of the X-ray spectra are detected in the AGN with $\log(N_{\rm H}/{\rm
  cm}^{-2})>21.5$. Local absorbers with $\log(N_{\rm H}/{\rm
  cm}^{-2})=21.5$ are
opaque for $E < 1$ keV and transparent for $E > 1$ keV. The
transition energy between the transparent and opaque regime linearly grows
with $ N_{\rm H}$, until $N_{\rm H} \sim 10^{24} \rm cm^{-2}$, where the
Compton Thick regime dominates.
We show in Fig. \ref{Fig:DistrLz_abs_unabs} the distribution of the
absorbed and unabsorbed subsamples just defined: the two
subsamples have a similar distribution in the luminosity-redshift
plane, so they are fair representations of the AGN with and
without absorption, with no other parameters playing an important role (see also Table
\ref{tproperties200}).

\item{Luminosity:} We separated high and low luminosity sources using a threshold $L=8\times 10^{43}$~erg~s$^{-1}$

\item{$z$:} We built a low-$z$ and a high-$z$ subsample with a threshold $z=1.005$

\item{$L-z$:} Initially we divided the full sample in three
  subsamples: (i) $z<0.76$ (hereafter low $z$-low $L$) (ii) $z>0.76$
  and $L<14.35\times10^{43}$~erg/s (hereafter high $z$-low $L$) (iii)
  $z>0.76$ and $L>14.35\times10^{43}$~erg/s (hereafter high $z$-high
  $L$). In the low $z$-low $L$ subsample there are only 3 sources
  above $L<14.35\times10^{43}$~erg/s (with $\sim$4000 total counts),
  so we have also used a second version of this subsample using only
  sources with $z<0.76$, $L<14.35\times10^{43}$~erg/s so that this
  subsample and the two high $z$ subsamples cover distinct $z$ and $L$
  intervals. All results are very similar and we have used the latter
  definition of distinct areas as our default.
 
\end{itemize}

   
\section{Analysis method}\label{method}

\subsection{Averaging method}

Our averaging method was presented for the first time in
\cite{corral2008} for the \emph{XMM-Newton} spectra of XMS-XWAS
surveys; it was then also used for the study of the XBS sample by
\cite{corral2011}. We adapted and tested it for the \emph{Chandra}
spectra.

After fitting the spectra as described in Sect.2.2, we read into
\texttt{XSPEC} again each un-binned, background-subtracted spectrum,
we applied the corresponding best fit model and we extracted the unfolded spectra taking into account the detector response (\texttt{eufspec} command in
\texttt{XSPEC}). In this step it is possible that some distortions to narrow spectral features were introduced, such as to the
Fe emission line. In
order to quantify this effect, we performed extensive simulations of
the continuum and of the Fe Line as described in Sect.3.2 and 3.3.
Once we obtained all the unfolded spectra, the following step was to
apply the corrections described in \cite{corral2008} that we can
summarize as follows:

\begin{enumerate}
      \item Correction for Galactic absorption

      \item Shift the spectra to a common redshift frame

      \item Re-normalize the spectra with respect to the integrated flux between 2 and 5 keV restframe
\end{enumerate}

Once we obtained the de-absorbed and re-normalized spectra at
rest-frame, we re-binned each spectrum.  For simplicity, we used the
same binning for all the average spectra, applying the bin definitions of the
absorbed sample: this binning was constructed in order to have at least 1000
net counts in each bin of the co-added spectrum. In order to maximize
 our ability to detect a narrow Fe line, we always centered one bin at 6.4
keV. We finally averaged the spectra using the
unweighted arithmetic average.

\begin{figure}
  \centering
  \includegraphics[angle=-90,width=8cm]{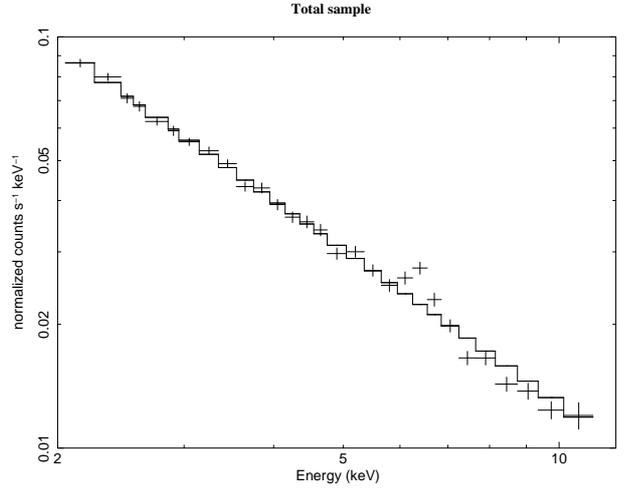}
  \caption{Average observed spectrum  of the full sample fitted with a powerlaw between 2 and 5 keV.}
  \label{Fig:All_po}
\end{figure}

We show in Fig. \ref{Fig:All_po} the average spectrum of the total
sample. If we fit the 2-5~keV range with a powerlaw and extrapolate it
to the full range, an excess can be clearly seen around the expected
position of the Fe emission line. However, any features found in the
average spectra must be carefully evaluated, as the averaging
procedure itself may introduce some distortions. For this reason, we
 computed the continuum using simulations, as described below.

\subsection{Simulations of the continuum}

Previous works, as for example \cite{Yaqoob} and \cite{corral2008}, have shown
that the unfolding and the averaging process can distort the shape of
the spectrum. Therefore, it is important to take this into account
before deriving conclusions from the spectral analysis of the average
spectrum.

In order to characterize the underlying continuum of our sample, we
made 100 simulations of each source using the best fit parameters of
the continuum model (see Sect. 2.2).  To each of these 100 simulated
samples we applied the same method as the one used for the observed
sample (spectral fitting, correcting for response, correcting for
Galactic absorption and z, rescaling and averaging). After this, we
represented our continuum with the median of the 100 averaged simulated
continua. We decided to use the median and not the arithmetic average
as it is a more robust estimate of a central value and in particular it is less sensitive to extreme values. Hereafter we
will call this the simulated continuum.

\begin{figure}
  \includegraphics[angle=-90,width=7cm]{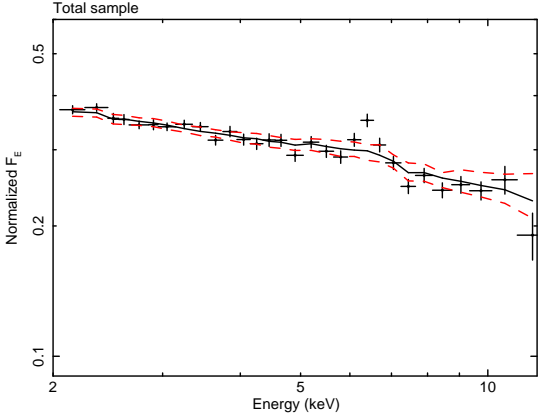}
  \caption{Average observed spectrum of the full sample (data points) with the average simulated continuum (continuous line) and the one sigma confidence limits (dashed lines).}
  \label{Fig:All_68}
\end{figure}

\begin{figure*}
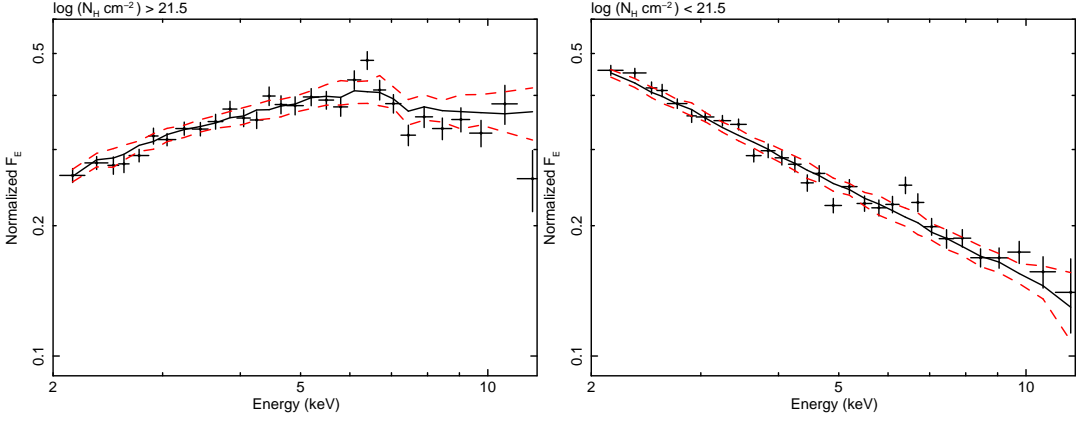

  \includegraphics[angle=-90,width=7cm]{./figures/mediana_ave_no_each_clip_abs_cts200_data_average_68_pl.ps}
  \includegraphics[angle=-90,width=7cm]{./figures/mediana_ave_no_each_clip_unabs_cts200_data_average_68_pl.ps}
  \caption{Average observed spectra (data points) of the absorbed (left) and unabsorbed (right) subsample with their average simulated continua (continuous line) and the one sigma confidence limits (dashed lines).}
  \label{Fig:Allnh_68}
\end{figure*}

\begin{figure*}
    \includegraphics[angle=-90,width=7cm]{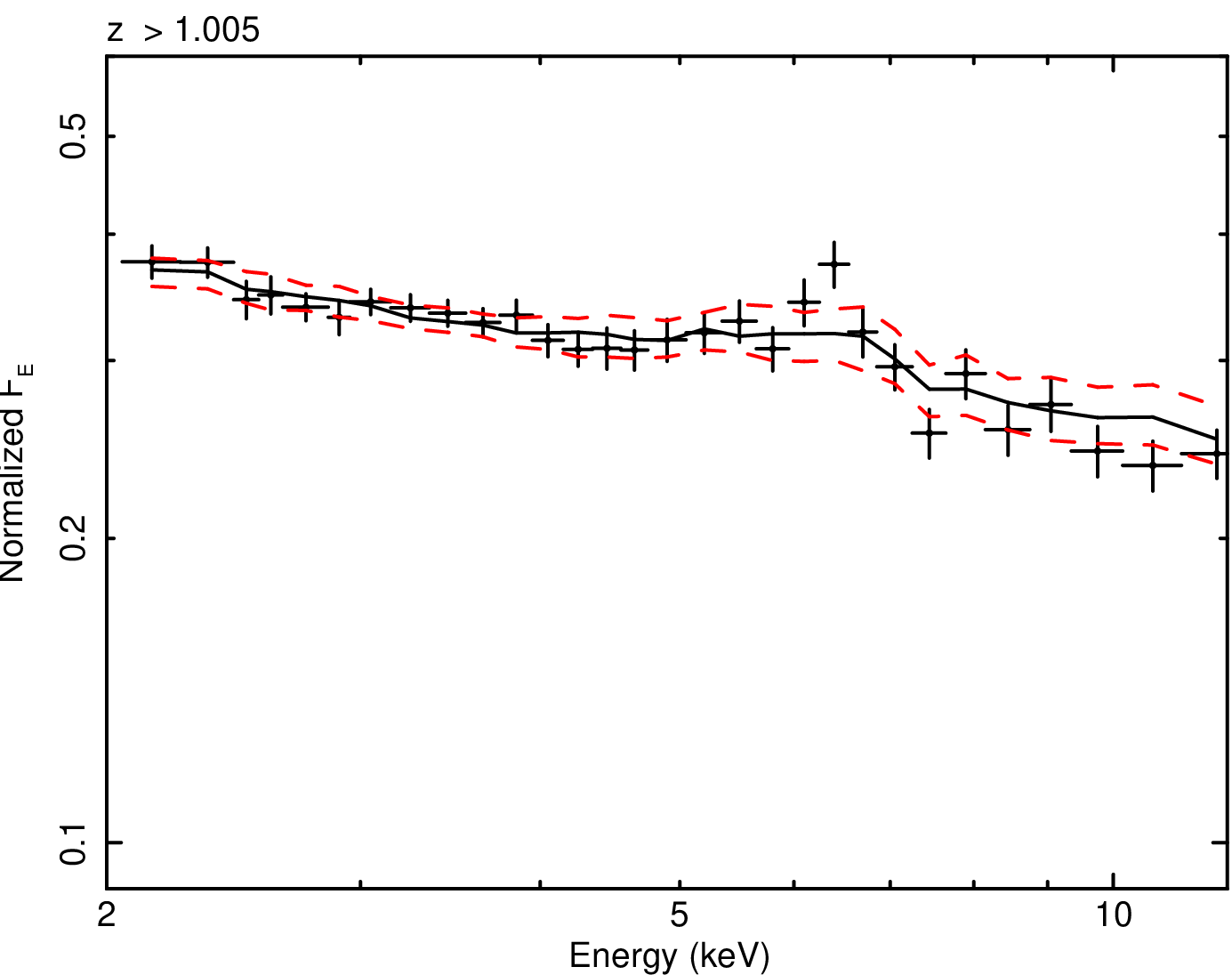}
    \includegraphics[angle=-90,width=7cm]{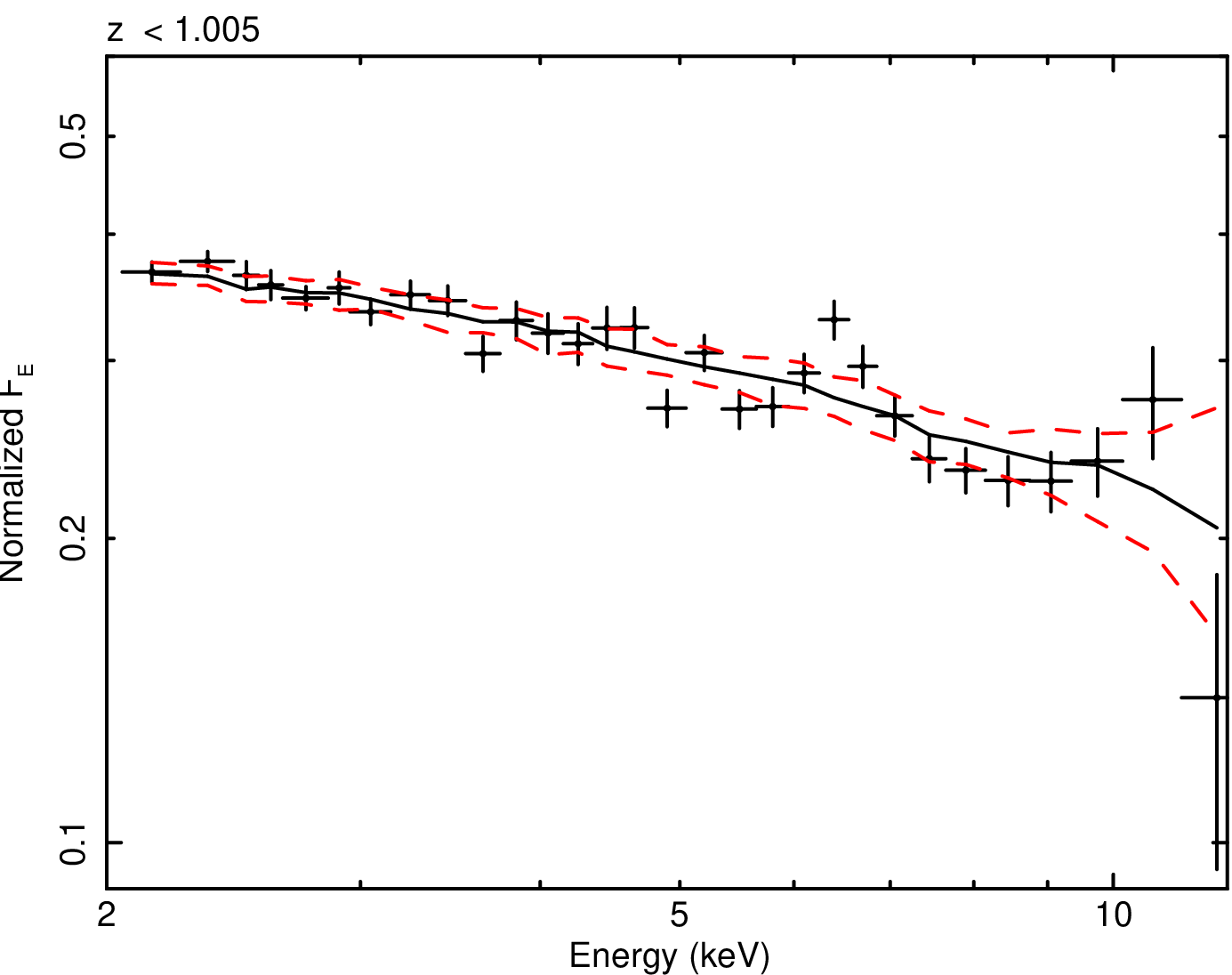}
    \caption{Average observed spectra (data points) of the high-$z$ (left) and low-$z$ (right) subsamples with their average simulated continua (continuous line) and the one sigma confidence limits (dashed lines).}
    \label{Fig:Allz_68}
  \end{figure*}

We represent in Figs. \ref{Fig:All_68}, \ref{Fig:Allnh_68},
\ref{Fig:Allz_68}, \ref{Fig:AllL_68}, and \ref{Fig:AllL_z_68} the average
observed spectra (dots with error bars), along with the simulated
continua (solid lines), and the 1 sigma confidence limits
(dashed lines). The latter correspond to the 16th and 84th element of
the sorted fluxes of the simulated spectra at each bin. Features
around the expected location of the Fe K line are conspicuous in all
the subsamples.

Before attempting to draw any conclusions from those features, we have
characterized the effect of our averaging method on narrow lines using
simulations, as described in the next section.

  
  \begin{figure*}
    \includegraphics[angle=-90,width=7cm]{./figures/mediana_ave_no_each_clip_lowLum_cts200_data_average_68_pl.ps}
    \includegraphics[angle=-90,width=7cm]{./figures/mediana_ave_no_each_clip_highLum_cts200_data_average_68_pl.ps}
    \caption{Average observed spectra (data points) of the high-$L$ (right) and low-$L$ (left) subsamples with their average simulated continua (continuous line) and the one sigma confidence limits (dashed lines).}
    \label{Fig:AllL_68}
  \end{figure*}

\begin{figure*}
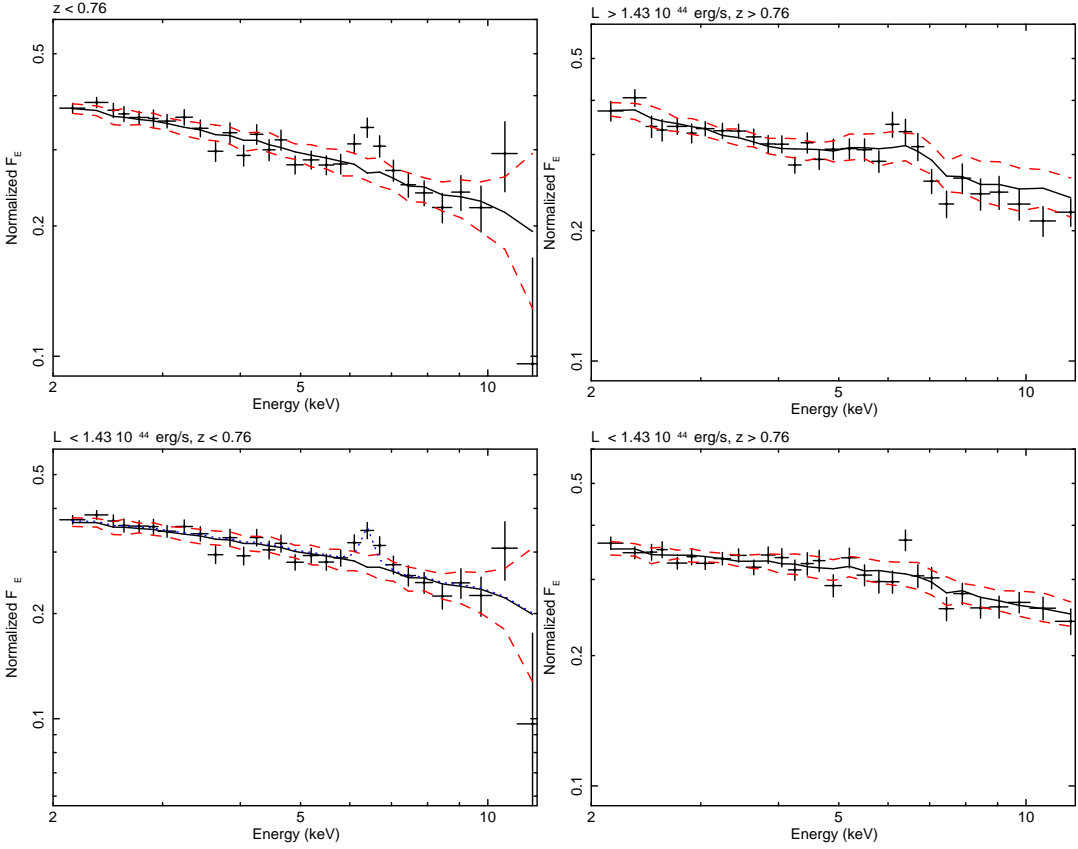

    \includegraphics[angle=-90,width=7cm]{./figures/mediana_ave_no_each_clip_lolo_cts200_data_average_68_pl.ps}
    \includegraphics[angle=-90,width=7cm]
{./figures/mediana_ave_no_each_clip_hihi_cts200_data_average_68_pl.ps}\\
\includegraphics[angle=-90,width=7cm]
{./figures/mediana_ave_no_each_clip_lolo__nohigh_lum_cts200_data_average_68_pl.ps}
  \includegraphics[angle=-90,width=7cm]
{./figures/mediana_ave_no_each_clip_hilo_cts200_data_average_68_pl.ps}
\caption{Average observed spectra (data points) of the $L-z$ subsamples with their average simulated continua (continuous line) and the one sigma confidence limits (dashed lines). For comparison, in the bottom-left panel we represent with a dotted line the result of adding to the simulated continuum a Gaussian centered at 6.4 keV, with $\sigma$=120 eV, as obtained from the simulations of the Fe line (Sect. 3.3).}
\label{Fig:AllL_z_68}
\end{figure*}

\subsection{Simulations of the Fe Line}

We performed simulations of unresolved Fe lines in order to study
how the spectral resolution of ACIS-I  and the averaging process widen the narrow spectral
features in X-ray spectra. 

We simulated high signal to noise spectra (one simulation for each
spectrum, without Poissonian noise) using a powerlaw with $\Gamma$
fixed at 1.9, with unit normalization, and a zero sigma Gaussian. The
input EW was 200 eV and energy fixed at 6.4 keV, at the
$z$ of each source. We made the correction for the response (using a
power law with Gamma fixed at 1.9) and we finally computed the
average spectrum with the same treatment applied to the real spectra.

We show in columns $\Gamma_{\mathrm simline}$, $\sigma_{\mathrm
    simline}$ and EW of Table \ref{tproperties200} the results of
  fitting a powerlaw plus a Gaussian to these simulations for some of
  the samples. The EW and powerlaw slope are recovered
  very well. The width of the line has become $\sigma\sim120$~eV.
Therefore, our resulting ``instrumental resolution'' is around
120 eV, i.e., any real detected feature around rest-frame 6.4
keV should be wider than this value and its actual width will be the convolution of its intrinsic width with this value, roughly added in quadrature (see Eq. 1 below).

We show in the bottom-left corner of Fig. \ref{Fig:AllL_z_68} the
  expected profile of an unresolved 6.4~keV line added to the
  simulated continuum for the corresponding subsample. In general, the
  features we observe around that energy appear to be compatible with
  such an unresolved line, except perhaps for the low $z$-low $L$,
  high $z$-high $L$ and high $L$ subsamples. We will quantify these
  qualitative impressions in Section \ref{specfits}.

\section{Results}

In this section we study the statistical significance of the Fe line
and its properties in several steps. We first developed a
model-independent way to calculate the significance of the line
detection (Sect. 4.1), then we studied again its significance and
characteristics fitting the average spectra using a single Gaussian to fit the Fe line (Sect. 4.2). In Sect. 4.3 we study
the dependence of the line on $z$ and X-ray luminosity.
Finally, in Sect. 4.4 we try complex line models with those
subsamples that show significant broad line profiles.

\subsection{Significance from the simulations}\label{analysis}

In order to estimate the significance of the excess observed in the Fe
K line region, we calculated the percentage of average
  simulated spectra (from the simulations of the continuum described
  in Sect. 3.2) having a flux in the Fe K region which is lower than the flux of the
observed spectrum. We should note that this method does not depend on
a modeling of the continuum (as in the calculation of the equivalent
width, in the next Section). The Fe line flux was
calculated in three regions centered at 6.4 keV, with half-widths of
0.1 keV, 0.2 keV, and 0.4 keV respectively. The significances of the
Fe line estimated in this way for the full sample and all subsamples are in Table
\ref{tfrac200}.

The narrowest interval was chosen at the limit of the
instrumental broadening, a significant excess found just in this range
and not in the other two would in principle correspond to a
detection of a narrow line. On the other hand, increasing the width
includes more flux from a putative broad line, but it also increases
the noise, hence a broad component would be more difficult to detect.
We use the $\pm0.2$~keV wide option (column $S_2$ in
  Table \ref{tfrac200}) as our fiducial value, since it maximizes the
  signal-to-noise ratio for unresolved and moderately broad lines. We
  can see that the line is strongly significant ($ > 98 $ \%) in the
  full sample and in almost all the subsamples, with the exception of
  the high $z$-high $L$ one. In general, the significance is higher
  for low $z$ and low $L$.

In all cases, the significance stays about the same or decreases
when $ \Delta E $ increases. This is either an effect of the
  expected lower signal-to-noise ratio for wider intervals, or might
  indicate the presence of a wider profile in the cases where the
  significance does not decrease noticeably.

\begin{table}[h]
\caption{\label{tfrac200}Significance of the Fe line estimated from the simulations for the full sample and the subsamples.}
\centering
\begin{tabular}{lccc}
\hline
Sample   &$S_1$  &$S_2$  &   $S_3$  \\
   &  (1)   &   (2) &  (3)   \\
\hline
\textbf{Total}  & 100 &  100 & 99  \\
\hline
\textbf{$\mathrm{log{(N_{\rm H})}}>21.5$ } & 99 & 99  & 98  \\

\textbf{$\mathrm{log{(N_{\rm H})}}<21.5$ } & 100 &  100 &  100 \\
\hline

\textbf{ $\mathrm L_{43} < 8   $}   & 100 &  100 &  100 \\

\textbf{ $\mathrm L_{43} > 8   $}   & 97 &   98  &  96 \\
\hline
\textbf{$\mathrm z <  1.005$}    &  100  &  100 &  100 \\

\textbf{$\mathrm z >  1.005$}    & 98 & 98  & 98  \\
\hline
\textbf {$\mathrm L_{43} < 14.3 $, $\mathrm z < 0.76 $} & 100 &  100 & 100  \\

\textbf {$\mathrm z < 0.76 $} & 100 &  100 &  100 \\

\textbf {$\mathrm L_{43} < 14.3 $, $\mathrm z > 0.76 $} & 99 &  98 &  91 \\

\textbf {$\mathrm L_{43} > 14.3 $, $\mathrm z > 0.76 $} & 88 & 88  &  86 \\

\end{tabular}
\tablefoot{ $L_{43}$: luminosity in units of $10^{43}$ erg s$^{-1}$. $ S_1 $, $ S_2 $, $ S_3 $: percentage of average simulated spectra (from the simulations of the continuum, see Sect. 3.2) with the line flux lower than the observed spectrum in the intervals: 6.3-6.5 keV (1); 6.2-6.6 keV (2); 6.0-6.8 keV (3).}
\end{table}

\subsection{Spectral fits on the full sample and subsamples}
\label{specfits}

After having made the above mentioned calculation of the significance
of the Fe line, we performed the spectral analysis. To do this, we
represented the continuum using the simulated continuum as a
\texttt{table model} in \texttt{XSPEC}. During the fits described below, we have aways left its normalization free to vary, and its value returned in \texttt{XSPEC} is always around one.
We added to the table model
the required components when we found residuals.  We used the goodness of fit
criterion and the confidence interval for a given parameter
corresponding to a $\mathrm \Delta\chi^2 = 2.71 $ (90 \% probability).
The fit results are shown in Table \ref{tresults200}. 

The $ \sigma $ we mention in the remainder of this work (including Table \ref{tresults200}), refers to the intrinsic width of the line, taking into
account the broadening introduced by our method

\begin{equation}
 \sigma^2 =
\sigma^2_{obs} - \sigma^2_{sim}
\end{equation}

\noindent where $ \sigma_{obs} $ is the total width returned by \texttt{XSPEC}  
and $\sigma_{sim} $ is the instrumental width obtained from the
simulations of an unresolved Fe K line (120 eV, see Sect. 3.3.). During the fits, we
forced the width of the Gaussian line to satisfy the condition:
$\sigma_{obs} \geq\sigma_{sim}$ (where we used the average value of $\sigma_{sim}$ reported in Table \ref{tproperties200}, that is approximately 120 eV).

We made the fits in several steps:
\begin{itemize}
\item Fixed $ \sigma$=0 and fixed centroid energy at 6.4 keV: this
  allows us to estimate the significance of the narrow component of a neutral
  Fe K line (first line for each sample in Table 3)
\item Fixed centroid energy at 6.4 keV and free $\sigma$: studies 
  the significance of a possible broad component and
  constrains its width (second line for each sample)
\item Free centroid energy and fixed $\sigma$=0: considers the
  presence of an ionized narrow Fe component and estimates its centroid
  energy (third line for each sample)
\item Free centroid energy and free $\sigma$: leaves all options open
  (fourth line for each sample)
\end{itemize}

We calculated the significance of the Gaussian with the $\Delta\chi^2$
corresponding to the fits with and without the line component, and we checked the corresponding probability using the incomplete $ \beta$ function (according to \cite{numericalrecipes}).  
The result is in Col. 4 in Table 3. In the same way, we estimated the significance of
allowing the width and/or the centroid energy to vary with respect to
the baseline narrow and neutral line.

\begin{figure}
  \centering
  \includegraphics[angle=-90,width=7cm]{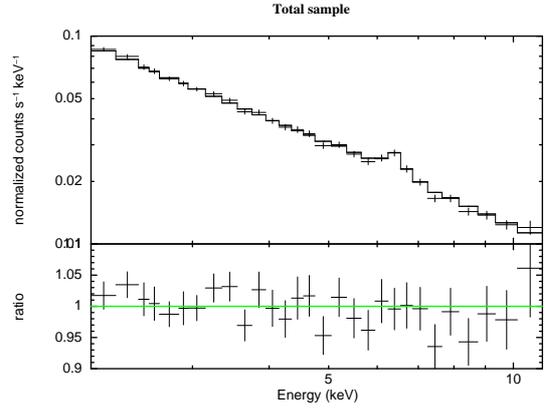}
  \caption{Fit of the average spectrum of the full sample using the
    simulated continuum model + Gaussian (with free $ \sigma $ and
    free centroid energy)}
  \label{Fig:AlltabGa}
\end{figure}

\begin{figure*}
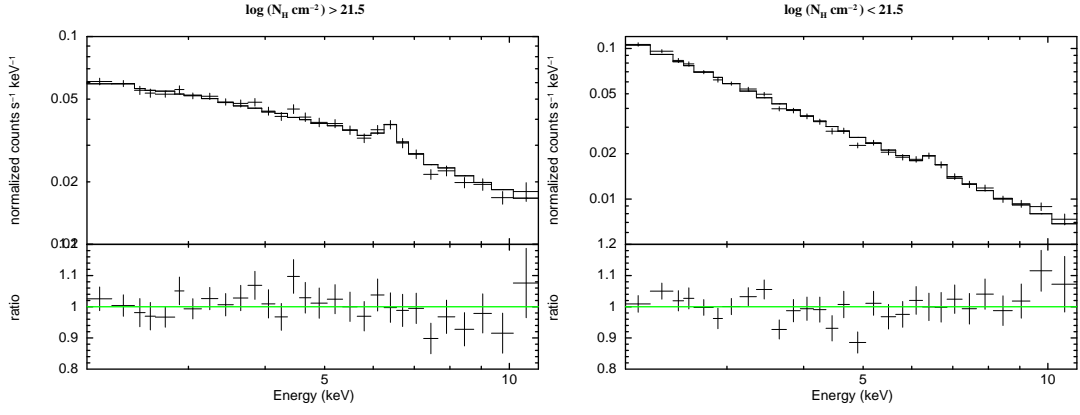

  \centering
  \includegraphics[angle=-90,width=7cm]{./figures/all_abs_sim_ga.ps}
  \includegraphics[angle=-90,width=7cm]{./figures/all_unabs_sim_ga.ps}
  \caption{Fit of the average spectrum of the absorbed (left) and unabsorbed (right)
    subsamples using the simulated continuum model + Gaussian (with
    free $ \sigma $ and free centroid energy)}
  \label{Fig:AbsTabGa}
\end{figure*}

\begin{figure*}
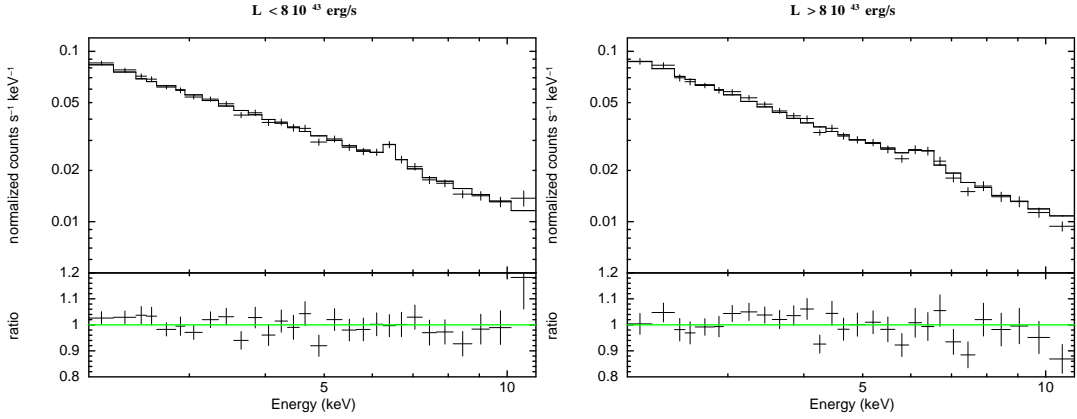


  \centering
  
  \includegraphics[angle=-90,width=7cm]{./figures/all_lowLum_sim_ga.ps}
  \includegraphics[angle=-90,width=7cm]{./figures/all_highLum_sim_ga.ps}
  \caption{Fit of the average spectrum of the low luminosity (left)
    and high luminosity (right) subsamples using the simulated
    continuum model + Gaussian (with free $ \sigma $ and free centroid
    energy)}
  \label{Fig:LbinsTabGa}
\end{figure*}

\begin{figure*}
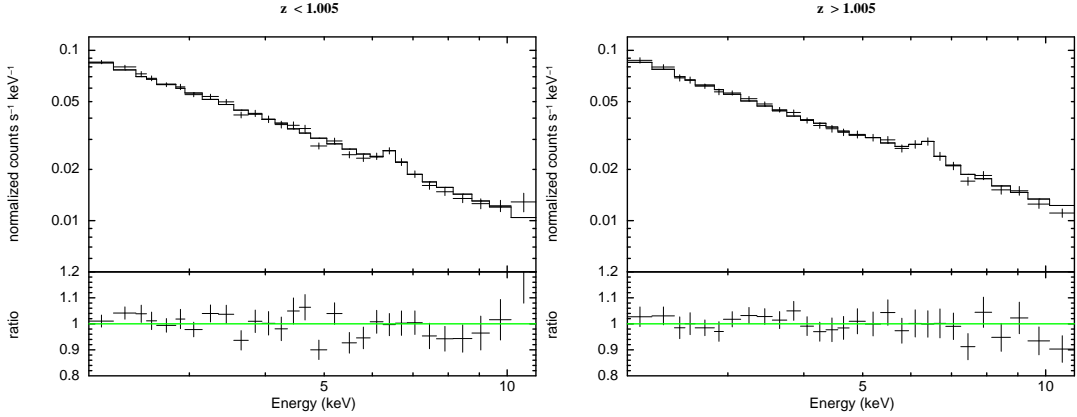


  \centering
  \includegraphics[angle=-90,width=7cm]{./figures/all_lowz_sim_ga.ps}
  \includegraphics[angle=-90,width=7cm]{./figures/all_highz_sim_ga.ps}
  \caption{Fit of the average spectrum of the low redshift (left) and
    high redshift (right) subsamples using the simulated continuum
    model + Gaussian (with free $ \sigma $ and free centroid energy)}
  \label{Fig:zbinsTabGa}
\end{figure*}

\begin{figure*}
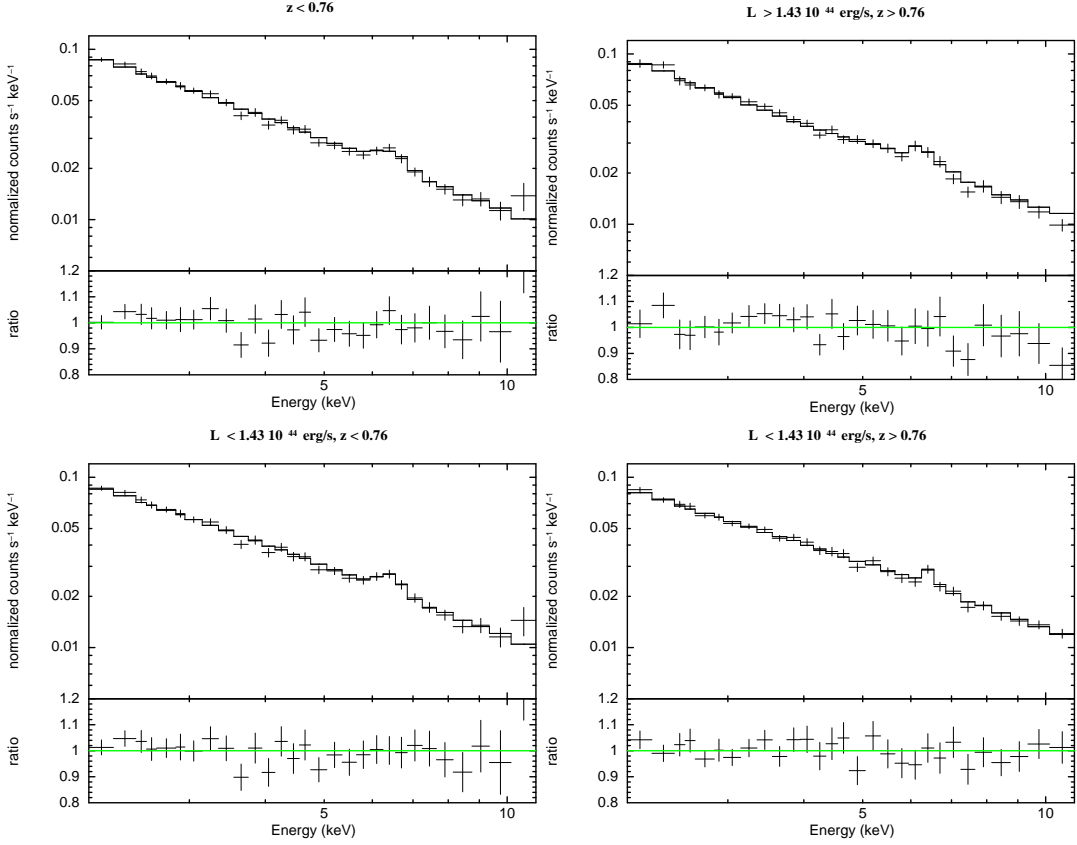
 

  \centering
  \includegraphics[angle=-90,width=7cm]{./figures/all_lolo_sim_ga.ps}
  \includegraphics[angle=-90,width=7cm]{./figures/all_hihi_sim_ga.ps}
  \includegraphics[angle=-90,width=7cm]{./figures/all_lolo_nohighLum_sim_ga.ps}
  \includegraphics[angle=-90,width=7cm]{./figures/all_hilo_sim_ga.ps}
  \caption{Fit of the average spectrum of the $L-z$ subsamples (low $L$-low $z$ subsamples in the left panels, low $L$- high $z$ subsamples in the right bottom panel, high $L$-high $z$ subsample in the right top panel) using the simulated continuum model + Gaussian (with free $ \sigma $ and free centroid energy)}
  \label{Fig:zLbinsTabga}
\end{figure*}

We summarize here the results for our (sub)samples (see Table
\ref{tresults200} and Figs. \ref{Fig:AlltabGa}, \ref{Fig:AbsTabGa},
\ref{Fig:LbinsTabGa}, \ref{Fig:zbinsTabGa}, and \ref{Fig:zLbinsTabga})
:

\begin{itemize}
\item \emph{Full sample: \\} The spectrum of the full sample fitted
  with the table model and the Gaussian with free energy and free
  $\sigma$ are shown in Fig.\ref{Fig:AlltabGa}. We can see that
  the model fits the spectrum well in the Fe line region.  The
  detected line is narrow. We do not observe any ionized Fe line
  component.
\item \emph{Absorbed and unabsorbed subsamples: \\} The spectra and
  fitted models are shown in Fig.\ref{Fig:AbsTabGa}. The continuum of
  the unabsorbed sample shows $ \Gamma\sim1.8$, a common value for the
  AGN (see Table 1), and is flatter in the absorbed sample, as
  expected.  The line is characterized by having a narrow profile for
  both subsamples. Only in the unabsorbed sample the centroid energy
  of the line might suggest a contribution from mildly ionized Fe.
  However, the centroid energy is consistent with 6.4 keV within the
  90 \% confidence level, and the significance of $\sigma>$0 from the $\Delta\chi^{2}$ is $<$ 90 \%.
\item \emph{High and low luminosity subsamples: \\} The spectra and
  fitted models are shown in Fig. \ref{Fig:LbinsTabGa}. The results of
  the fits suggest a more intense Fe line in the low-$L$ subsample.
  We discuss the dependence of the EW on redshift and
  luminosity more in detail in the next Section.  The detected line is
  narrow, and there is no significant ionized Fe contribution. 
\item \emph{High and low redshift subsamples: \\} The spectra and
  fitted models are shown in Fig. \ref{Fig:zbinsTabGa}.  The Fe line
  is more significant at low redshift than at high redshift and this
  can reflect the same trend just found with the luminosity, since
  sources with higher $L$ are preferentially found at higher $z$.  The
  line profile is narrow in both cases. The fits with the centroid
  energy free shows the presence of a mildly ionized Fe contribution
  in the low redshift sample, but 6.4 keV is within the 90 \%
  confidence interval.
\item \emph{$L$-$z$ subsamples: \\} The strongest
  line significance is found at low $z$ and low $L$, then decreasing
  with both redshift and luminosity.  The line profile is broad only in one
  case, in the low redshift- low luminosity sample ($>$95.4 \%
  probability). The line profile looks symmetric (see Figs. 14
  and \ref{Fig:zLbinsTabga}, two left panels). This suggests reflection
  from regions in newtonian movement in the accretion disc, although we can not exclude the relativistic profile. Moreover, in the top-left panel there seems to be
  an excess in the narrow bin around 6.4~keV, which might come from an
  additional narrow component. We discuss in more detail the shape of
  this feature in Section \ref{broadLines}.  In the other two
  subsamples, the detected profile of the Fe line is narrow.  We did
  not find any ionized Fe component.
\end{itemize}

At high energies ($\sim $ 10 keV), we observed in some spectra an
excess that can be interpreted as part of the Compton reflection. The
paucity of counts at these energies does not allow us to make a more
accurate assessment of this continuum component.  An absorption feature was
also detected, to some extent, in some of our average
spectra around 7-8 keV: this can be explained as the Fe edge
commonly found along the Compton reflection component.

\subsection{Dependence on redshift and luminosity}

\begin{figure*}
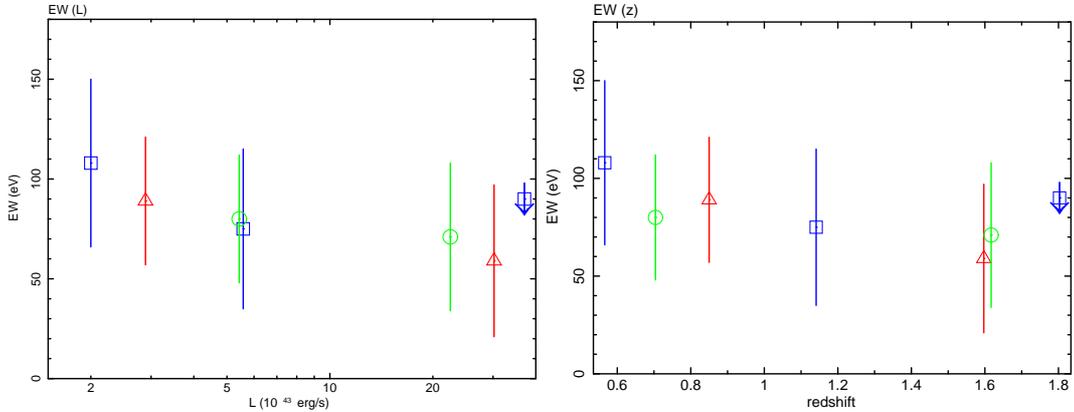

   \centering
   \includegraphics[angle=-90,width=7cm]{./figures/eqw_L.ps}
   \includegraphics[angle=-90,width=7cm]{./figures/eqw_z.ps}
   \caption{Dependence of unresolved 6.4~keV Fe line equivalent
     width with the average luminosity (left) and the average redshift
     (right) of the subsamples (first row of Table 3). Squares: $L-z$
     subsamples; circles: redshift subsamples; triangles: luminosity
     subsamples. We represent upper limits with a
       down-pointing arrow. Each set of points with the same symbol
       are statistically independent, but not so across symbols.  }
         \label{Fig:eqw_L}
   \end{figure*}

As our results seem to suggest a dependence of the Fe line EW
on redshift and on luminosity, for low $z$ and low luminosity
  AGN, we checked this result in more detail to understand it better. The
distribution of the line EW with the average $z$ and the
average luminosity of the subsamples are shown in Fig.
\ref{Fig:eqw_L}.  In those figures, the same symbols refer to
  statistically independent subsamples.  On the contrary, subsamples
  with different symbols are not statistically independent and there
  will be significant overlapping among the sources included in each
  subsample. We have used the fits corresponding to an unresolved line
  fixed at 6.4~keV (first line for each subsample in Table 3).

At first view the EW seems to decrease both with increasing $z$ and X-ray
  luminosity. However, looking for example at the blue squares
  (corresponding to independent $L-z$ subsamples), it is clear that a
  constant EW is consistent with the data points the upper limits
  and, hence, the trend, albeit suggestive, is not statistically significant.

\subsection{Analysis of broad lines}
\label{broadLines}

We detected with strong significance narrow Fe K lines in all samples.
Additionally, in the low $L$-low $z$ subsample, the Fe line appears significanlty broadened with a significance above two sigma). 
We investigate here whether relativistic profiles are
  better fits than a simplistic broad Gaussian.

We added to the simulated continuum the \texttt{diskline} model
in \texttt{XSPEC} which describes a relativistically broadened emission line for an accretion disc around a Shwartzschild Black Hole.  We fixed $R_{out}$, the external radius of the
accretion disc, to $1000\,R_{g}$. We left $R_{in}$, the inner radius,
free to vary, obtaining an upper limit for it: $<142\,R_g $.  As
mentioned in Sect. 1, the inner radius of the accretion disc is
expected to be lower ($6\,R_g $ for a non-rotating black hole and
$1.23\,R_g $ for a maximally-rotating one).  The EW is
consistent with the one obtained in the fit with the Gaussian (see
Table 3): EW=$152^{+70}_{-70}$~eV. The significance of adding an Fe
line with \texttt{diskline} to the continuum is $>99.99$\%, again
consistent with the case of the Gaussian (Table 3). The fit gives
$\chi^2/$dof=20.53/29, to be compared with $\chi^2$/dof=19.07/25 for
the Gaussian: the fit is not better with more free parameters.
Summarizing, the diskline fits the Fe K line as well as the Gaussian:
we do not significantly detect a definite relativistic profile in the line, although it cannot be excluded.

Alternatively, the profile of the line could be characterized as the
sum of both a broad and a narrow components since, in principle, we
would expect a contribution from both a torus (narrow component) and
disk (broad component) reflections.

We have assessed the significance of this possibility by making a final test in the low $L$- low $z$ sample: we added a broad Gaussian to
the narrow Gaussian (i.e. combining models in the first and second
line of Table 3). We fixed the centroid energies of the two Gaussians
at 6.4 keV and we fixed the $\sigma$ to zero in the narrow Gaussian
and to 240 eV (the value in the second line in Table 3) in the second broad
Gaussian.

We calculated the EW of the narrow and broad components, obtaining
$<84$ eV and $158^{+78}_{-78}$~eV, respectively.  This fit gives
$\chi^2$/dof=19.15/26, to be compared with $\chi^2$/dof=23.70/27 of
the fit with only narrow Gaussian (in Table 3). Hence, the
significance of adding the broad component ($\Delta\chi^{2}$=4.55 and
$\Delta\nu$=1) is $>95.4$\% but $<99$\%, calculated as mentioned in Sect. 4.2. We conclude that a double Gaussian is a better fit,
  but only at the $2\sigma$ level.

\section{Conclusions}\label{conclusions}

We have studied the average spectrum of a sample of 123 AGN with
more than 200 net counts selected from the \emph{Chandra} AEGIS, CDF-N
and CDF-S surveys, covering $\mathrm 0 < z < 3.5$ and totalling
  $>$70000~cts.  Compared to similar \emph{XMM-Newton} and
  \emph{Chandra} studies of deep fields, we sample better the
  luminosity-redshift plane at $z\leq3$ (including higher
  luminosities) and use deeper CDF-S data. With respect to shallower
  wider surveys using \emph{XMM-Newton}, we reach about an order of
  magnitude lower luminosities over the same redshift range.

To improve the signal-to-noise ratio and to study the average
properties of the sample, we computed its average spectrum,
adapting the method developed by \cite{corral2008}. We constructed the
continuum model with simulations, in order to accurately analyze our
resulting spectra. We also assessed the effect of our averaging
  method on unresolved features around 6.4~keV, to obtain the intrinsic
  widths of any putative lines.

We repeated the averaging procedure for the subsamples defined in
  two intervals of column density, 2-10~keV X-ray luminosity,
  redshift, and in three 2D intervals in the luminosity-redshift
  ($L-z$) plane (see Table~\ref{tproperties200}).

  We have estimated the significance of the presence of narrow and
  broad features around 6.4~keV in a model-independent way (Table 2),
  finding that narrow features are significant at $\geq$98\%, except
  for the high-$L$, high-$z$ subsamples, where it is lower. Broad features would be harder
  to detect and are most significant at low $L$ and low $z$.

Analyzing the spectra with \texttt{XSPEC} using the
  simulated continuum and modelling the Fe line as a Gaussian, we have
  found strong evidence of a narrow Fe line at high significance in
  our full sample and in most of our subsamples, in particular (Table
  \ref{tresults200}):

   \begin{itemize}
   \item We detected a strong (at $> 99.73$ \%) narrow Fe line
       with EW$=85\pm35$~eV and $73\pm32$~eV in the absorbed
       and unabsorbed subsamples, respectively (defined as having
       intrinsic column density above and below $10^{21.5}$~cm$^{-2}$,
       respectively), with a hint of higher central energy
       $\sim$6.5~keV in the unabsorbed case, but still compatible with
       6.4~keV at 90\%.
   \item Segregating the sources purely on luminosity, we found
       that the average spectrum in the low-$L$
       ($L<8\times10^{43}$~erg~s$^{-1}$) subsample shows a significant
       and strong narrow Fe line (EW$=89\pm32$~eV at $>$99.99\%).
       The significance and strength of the high-$L$ are clearly
       smaller: EW$=59\pm38$~eV at 95.4\%.
   \item Separating instead at $z=1.005$, the narrow Fe line is again
     stronger in the low-$z$ subsample (EW$=80\pm32$~eV at $>$99.99\%)
     than in the high-$z$ subsample (EW$=71\pm37$~eV at $>$99.73\%)
     but the difference is not significant
   \item Defining distinct areas in the $L-z$ plane (the division
     point was $L=14.3\times10^{43}$~erg~s$^{-1}$,$z=0.76$), the most
     significant line is found in the low $L$-low $z$ subsample
     (EW$=108\pm42$~eV at $>99.99$\%), with a hint of a broad
     component ($\sigma=240\pm90$~eV at $>2\sigma$ significance). The
     low $L$-high $z$ subsample also shows a somewhat weaker narrow
     line (EW$=75\pm40$~eV at $\sim99.7$\%). The Fe line detection is
     at $<90$\% significance in the high $L$-high $z$ subsample
     (EW$<90$~eV)
\end{itemize}

We studied the trend of the EW for the narrow Fe line with redshift and
  luminosity of our subsamples. We did not find a significant
  dependence: in Fig. \ref{Fig:eqw_L} that trend is not statistically
  more significant than a constant. In order to disentangle the
  effects of the two parameters on the line detection, surveys of AGN
  covering a wider area in the $L-z$ plane are needed. Since the Fe
  line is detected more significantly at low-$L$ and low-$z$, samples
  of the local Universe are particularly suited to this end.

We have also investigated whether more sophisticated models for the
  line profile would provide a better fit, concentrating on the low
  $L$-low $z$ subsample, since it is the only one where a broad line
  may have been detected. A \texttt{diskline} relativistic profile
  provides a worse fit than a broad Gaussian, with more parameters.
  Additionally, the inner radius of the disk would be too large to
  produce significant relativistic effects.  Therefore, we do not find
  evidence for a relativistic profile in any of our average spectra.

However, as has been pointed out in \cite{Guainazzi}, in the average
  spectra the broad component of the observed line comes from a
  maximum $\sim50$\% of the sample: the line in an average spectrum
  cannot have a relativistic profile as definite as the one that can
  be in principle observed in a single, good quality spectrum. In
  particular, there can be a contribution from partially ionized Fe
  that produces narrow lines at energies $> $ 6.4 keV, up to $\sim$7
  keV if the Fe is completely ionized. Our observed
  $\sim$symmetric Gaussian profile may hence be a combination of
  many mildly ionized Fe lines with a red wing, perhaps from
  relativistic effects.  As mentioned in Sect. 1, a similar result
  has been recently found in the stacking of the X-ray spectra of the
  COSMOS sample (\cite{iwasawa2011}).

Allowing for both a narrow (from neutral material far away from the
  central source, e.g. the putative torus) and a broad Gaussian (as
  just discussed) components provides a better fit than a single
  narrow or broad Gaussian, but with a modest significance $<99$\%.

Further spectral averaging studies with higher statistics
  covering a wider range of source properties would allow a more
  detailed characterization of the Fe feature in the X-ray spectra
  of AGN, and whether it depends on cosmic time, intrinsic brightness,
  amount of surrounding material, black hole mass, accretion
  rate, and other physical parameters, helping to constrain the physical properties of AGN
  throughout the history of the Universe.

\newpage
 \begin{table*}[h]\label{tresults200} 
 \caption{Results of fits of the average spectrum of the full sample and its subsamples.}
 \begin{tabular}{lllllllllll}
 Sample     &    $\chi^2$/dof(s) & $\chi^2$/dof(g) &
 $P_1 $&   $P_2 $ &  E  & $\sigma$ & EW  & $\langle z \rangle$ & $\langle L_{43} \rangle$ & $N_{h,22}$ \\ 
      &     &  & \%
 & \%  &  keV  & eV & eV  &  & $10^{43} $erg s$^{-1}$ & $10^{22}cm^{-2}$ \\ 
   (1)    &  (2)   &  (3)  &    (4)  &   (5)  &   (6)   &  (7)  &  (8)   &  (9)  &  (10)  &  (11)  \\ 
  \hline
\textbf{Total}   & 47.50/28  & 23.74/27   & $> 99.99 $  &  & 6.4  & 0  & $74 \pm 24$  & 1.15  & 13.80  & 1.75   \\ 
   &   & 21.98/26  &  $> 99.99 $ &  $<90$ & 6.4  & $<138$  & $94 \pm 30$  &     &   &    \\
 
     &    &  23.74/26    &  $> 99.99 $ & $<90 $  & $6.40 \pm 0.10 $  &  0  & $73 \pm 25$  &     &  &     \\
 
     &   & 21.98/25    &  $> 99.99 $   & $<90 $ & $6.40 \pm 0.10$  & $<123$   & $92 \pm 28$ &    &   &           \\

 \hline 

\textbf{$\mathrm{log{(N_{\rm H})}}>21.5$ }    & 35.61/28 & 21.66/27  & $> 99.73 $    
& $ $ & 6.4  & 0  & $85 \pm 35$  &  1.15   & 14.11  &   3.93 \\  
 
    &  & 21.15/26   & $> 99.73 $   
& $<90 $ & 6.4  & $<123$  & $100 \pm 45$  &    &   &     \\
 
    &   &   21.42/26   & $> 99.73 $     &  $<90 $ & $6.28 \pm 0.02$  &  0  & $92 \pm 41 $  &    &   &      \\

   &   & 21.22/25  & $> 99.73 $ & $<90 $  &  $6.40 \pm 0.30$  & $<151$   & $91\pm 39$ &    &   &  \\

\hline

\textbf{$\mathrm{log{(N_{\rm H})}}<21.5$}   & 53.03/28  & 39.38/27   & $> 99.73 $    & $ $ & 6.4  & 0  & $73 \pm 32$  & 1.14  & 13.55  & 0.045   \\   
 
    &  & 38.18/26  &    $> 99.73 $  & $ <90 $  & 6.4  & $<189$  & $100 \pm 43$  &   &   &      \\  
 
    &   & 37.41/26      &  $> 99.73 $   & $<90 $   & $6.50 \pm 0.10 $  &  0  & $85 \pm 36$  &   &  &        \\
 
    &   & 37.15/25   & $> 99.73 $    & $<90 $ & $6.50 \pm 0.10 $  & $<59.6$   & $87 \pm 36$ &      &   &         \\
\hline

 \textbf{ $\mathrm L_{43} < 8   $}    & 44.26/28 & 23.96/27    &   $> 99.99 $   &  $ $  & 6.4  & 0 & $89 \pm 32$     
 & 0.85    & 2.89  & 1.64  \\   

    &    & 23.70/26     &    $> 99.99 $   
  & $<90 $      & 6.4  & $<72$ &  $97 \pm 35$   
  &     &   &   \\   

     &  & 23.96/26    &    $> 99.99 $    
 & $<90 $ & $6.4\pm 0.1$   &  0  &  $89 \pm 32$    &     &   &   \\   

    &  & 23.56/25    &  $> 99.73 $     
 &  $<90 $ & $6.4\pm 0.9$  & $<68$  &   $97 \pm 34$  &     &   &   \\  
\hline

\textbf{$\mathrm L_{43} > 8   $}       & 42.59/28   & 36.42/27    &     $> 95.4 $   
&  $ $   & 6.4  & 0  &  $59 \pm 38$   &  1.597    & 	30.20		 & 1.91  \\ 

      &   & 36.41/26    &   $\sim 95.4 $      & $<90 $  & 6.4  & $<7$  &   $58 \pm 39 $  &      & 			 &   \\

        &  & 36.42/26    &     $> 90.0 $     & $<90 $ & $6.4 \pm 0.2$  & 0  &$58\pm 39$ &      & 			 &   \\

        &  & 34.86/25    &   $> 90.0 $        &  $<90 $  & $6.30 \pm 0.20 $  &$<7$  &   $75 \pm 45$   &      & 		 &  \\
\hline

\textbf{ $\mathrm z <  1.005$}         & 52.34/28 & 35.26/27    &    $> 99.99 $    & $ $  & 6.4  &  0  &   $80 \pm 32$   &  0.704    & 	5.43		 &      1.28  \\

        &  & 34.21/26   &   $> 99.73 $     & $<90 $   & 6.4   &$<100$  &  $94 \pm 37$   &      & 			 &        \\ 

         &  & 33.69/26    &  $> 99.99 $       & $<90 $ & $6.50 \pm 0.10$  &   0 &   $92 \pm 35 $   &      & 			 &        \\

            &   & 33.28/25    &    $> 99.73 $    &   $<90 $    & $6.45 \pm 0.10$  &$<101$  &   $99 \pm 37$ &      & 			 &        \\
\hline

\textbf{ $\mathrm z >  1.005$}       & 29.56/28   & 19.77/27    &    $> 99.73 $      & $ $   &  6.4  & 0   &   $71 \pm 37$  &  1.617    & 		22.5 & 2.24  \\  

        &    & 19.45/26    &    $> 99.73 $      &   $<90 $   &  6.4  & $<120$ & $86 \pm 44$    &      & 		 &  \\  

         &    & 19.77/26    &    $> 99.73 $      & $<90 $ & $6.40 \pm 0.15$  & 0  & $71 \pm 37$ &   & 		 &   \\

         &    & 18.66/25    &   $> 99.73 $       & $<90 $ & $6.30 \pm 0.10$  & $<76$ &   $85 \pm 43$   &      & 	 &   \\
\hline

\textbf {$\mathrm L_{43} < 14.3 $, $\mathrm z < 0.76 $}     & 42.01/28 & 23.70/27    &      $> 99.99 $   &  $ $ & 6.4  & 0 &   $108 \pm 42$    &  0.566   & 		
2.	 &  1.254 \\

      &  & 19.23/26    &      $> 99.99 $  &   $>95.4 $ & 6.4  & $240 \pm 90$ &  $178 \pm 63$    &     & 		
	 &   \\

       &   & 23.70/26    &      $> 99.73 $ &   $<90 $ &  $6.40 \pm 0.15$  & 0 &  $ 108 \pm 42$     &     & 		
	 &   \\

      &  & 19.07/25    &     
$> 99.99 $ &   $>90 $ &  $6.42 \pm 0.12$  & $216 \pm 50$ &   $172 \pm 60$  &     & 			 &   \\
\hline

\textbf { $\mathrm z <  0.76$}       & 43.28/28 & 23.04/27   &   $> 99.99 $   & $ $   & 6.4  & 0 &   $108 \pm 40$    &  0.573    & 	
3.23	 & 1.173  \\

      &   & 19.04/26    &   $> 99.99 $   & $\sim 95.4 $ & 6.4  & $157 \pm 30$ &   $154 \pm 50$        &      & 	
	 &   \\

       &  & 23.04/26   &    $> 99.99 $  & $<90 $ &   $6.40 \pm 0.15$ & 0  & $109 \pm 39$ &      & 	
	 &   \\
        
      &   & 18.72/25    &   $> 99.99 $   &  $<90 $ &  $6.43 \pm 0.11$   & $215 \pm 50$ & $170 \pm 56$       &      & 	
	 &   \\       
\hline

\hline
\textbf { $\mathrm L_{43} < 14.3  $, $z >  0.76$}      & 25.97/28  & 16.99/27     &      $ \sim 99.7 $ & $ $ &  6.4  & 
0 &   $75 \pm 40$   &  1.141    & 	5.58	 &  1.900 \\

    &  & 16.99/26    &   
 $ > 95.4 $ & $ <90$  &  6.4  & $<4$  & $75 \pm 40$  &      & 		 &   \\     

     &  & 16.99/26   &       $ > 95.4 $& $<90 $   & $6.40 \pm 0.10$   &  0   & $75 \pm 40$    &      & 		 &   \\      

     &  & 16.83/25    &       $ > 95.4 $&  $<90 $ &  $6.43 \pm 0.10$   & $<5$ &  $76 \pm 42$    &   & 		 &   \\
 \hline 

\textbf { $\mathrm L_{43} > 14.3 $, $z >  0.76$}     &  31.85/28  & 29.56/27     &      $ < 90.0 $ &  $ $ &  6.4  & 
0 &  $<90$    & 1.803     & 37.05		 &  2.187 \\

        &  & 29.56/26    &      $ < 90.0 $ &  $<90 $ &  6.4  & $<5$ &    $<91$        &     & 		 &   \\

    &  & 29.09/26    &      $ < 90.0 $ & $<90 $   &  $6.05 ^{+0.60}_{-0.05}$ & 0  & $50 \pm 47$   &      & 		 &   \\

     &  & 26.72/25    &      
 $ < 90.0 $ &  $<90 $ &  $6.22 \pm 0.15$   & $<5$ &  $73 \pm 53$   &      & 		 &   \\       
\hline
 
 \end{tabular}
 \tablefoot{$L_{43}$: luminosity in units of $10^{43}$ erg s$^{-1}$. Columns: (1): sample; (2): $\chi^2$/dof of the fit with simulated continuum model; (3): $\chi^2$/dof of the fit with the same continuum and the Gaussian; (4): probability $P(\Delta \chi^2, \Delta \nu)s,g $ (of adding the Gaussian to the model, see text); (5): probability $P(\Delta \chi^2, \Delta \nu)g,g_0 $  (of leaving the parameter of the Gaussian free, see text); (6): central energy of the Gaussian; (7):  $\sigma$ after subtracting the intrinsic sigma obtained in the simulations of the Fe line; (8): EW of the Gaussian; (9): average redshift of the sample; (10): average luminosity of the sample in $ (10^{43})$ erg s$^{-1}$; (11): average column density of the sample in $10^{22}cm^{-2}$. }
 \end{table*}

\begin{acknowledgements}
      Financial support for this work was provided by the Spanish Ministry of Science and Innovation through the grants \emph{AYA2009-08059} and \emph{AYA2010-21490-C02-01}.\\
      The authors acknowledge the computer resources, technical expertise and assistance
provided by the Spanish Supercomputing Network (RES) node (Altamira) at Universidad de
Cantabria in Santander; we thank especially Luis Cabellos for the technical user support.\\
We thank prof. Andrea Comastri and prof. Giorgio Matt, for useful comments.\\
This research has made use of data obtained from the Chandra Data Archive and the Chandra Source Catalog, and software provided by the Chandra X-ray Center (CXC) in the application of the package CIAO. \\
This study makes use of data from AEGIS, a multiwavelength sky survey conducted with the Chandra, GALEX, Hubble, Keck, CFHT, MMT, Subaru, Palomar, Spitzer, VLA, and other telescopes and supported in part by the NSF, NASA, and the STFC. \\
This research has made use of NASA's Astrophysics Data System.
\end{acknowledgements}


\appendix
\section{Results with different definitions of the sample}\label{appendix_results}

We assessed the robustness of our results by changing the
signal-to-noise ratio that characterizes our sample.  To do that, we
repeated the analysis for all the sources with more than 50 and 100
counts in 2 - 12 keV.  We excluded from these samples the spectra with
the lowest continuum flux between 2 - 5 keV (a bias during the
normalization process can be introduced, see Sect.3.1 for the details
for method): CDFS$_{-}$227 (CDF-S source with RA: 53.082, DEC:
-27.690), CDFN$_{-}$405 (CDF-N source with RA: 189.431, DEC: 62.177),
EGS1$_{-}$003 (AEGIS source with RA: 215.76, DEC: 53.45).  After
having defined the samples, we redefined the subsamples of
intrinsic column density, luminosity and redshift, as described in
Sect. 2.3. The results are in Table \ref{tproperties50} and
\ref{tproperties100}. We can see from the last column in the tables that
the Fe line significance grows with the signal-to-noise ratio of the
sample.  After having applied the same analysis method used for
  the default $>200$~cts sample, we repeated the analysis in
  \texttt{XSPEC} for the $>50$~cts and $>100$~cts sample, finding
  consistent results.

We have made the simulations of an unresolved 6.4 keV line for
  the $>50$~cts sample, obtaining results compatible with that of the
  default $>200$~cts sample.

\begin{table*}[h]
\caption{\label{tproperties50}Properties of the sample composed by all the sources with more than 50 counts.  }
 \centering
\begin{tabular}{lrrrrrrl}
\hline\hline
Sample &  $N$  &   $N_{2-12}$  & $N_{5-8}$   & $\langle z \rangle$ & $\langle L_{43} \rangle $  & $\langle N_{h,22} \rangle$     & Significance  \\
 &    &     &    &  & $ 10^{43} $ erg s$^{-1}$  & $10^{22} \times$ cm$^{-2}$      &   \\
 (1)  &  (2)  &  (3)   &  (4)  & (5) &  (6) &  (7)   & (8)  \\

\hline
 \textbf{Full}     & 347  & 93385  & 21730  &   1.20 &  8.60  & 3.08  &   85   \\
 \hline
\textbf{$\mathrm{log{(N_{\rm H})}}>21.5$}    &   177   & 40174      &    10944  &     1.14  &  8.00  & 6.04 & 85 \\
\textbf{$\mathrm{log{(N_{\rm H})}}<21.5$}   &   170  & 53211  & 10786  &     1.28 &  9.50   &  0.04 & 72 \\
\hline
\textbf{$\mathrm L_{43}<6 $ }        &  234  &  46546 & 11021  &     0.90 &  1.60 &  3.09 &  89\\

\textbf{ $\mathrm L_{43}>6 $ }      & 113   & 46840 &  10709 &    1.80 &   23.40  &  3.07 & 74 \\
\hline
\textbf{$\mathrm z<1$ }       &  184  & 46672 & 10530  &    0.65 &  2.60 &  3.08 &  83\\

\textbf{$\mathrm z>1$ }      &  163  &  46713   & 11199  &    1.83 &  15.60 & 3.08 & 85 \\
\hline
\textbf{$\mathrm  z < 0.76$}      &  120  & 31537 & 7511  &    0.53 &   1.80 & 3.75  & 81  \\

\textbf{ $\mathrm  L_{43} < 8.4 $, $\mathrm  z < 0.76$ }  &  116  &   26743  & 6614  &     0.52   &   1.20 & 3.35   & 82 \\

\textbf{$\mathrm  L_{43} < 8.4 $, $\mathrm  z > 0.76$ }       &  157  & 31862 &  6866 &     1.34  & 1.70 
  &  2.47 & 94 \\

\textbf{$\mathrm  L_{43} > 8.4$, $\mathrm  z > 0.76$ }       &  70  &  29985 & 7353  &     2.08 &   32.10     &  3.32 & 77 \\

\hline
\end{tabular}
\tablefoot{$L_{43}$: luminosity in units of $10^{43}$ erg s$^{-1}$. Columns: (1): (Sub)Sample; (2): Number of sources; (3): Number of counts in 2 -12 keV; (4): Number of counts in 5-8 keV; (5): average redshift;  (6): average luminosity in $10^{43}$ erg s$^{-1}$; (7): average column density of the local absorber in $10^{22}\cdot$ cm$^{-2}$ ; (8): significance of the Fe line calculated as the number of average simulated spectra with a lower flux than the flux of the average observed spectrum (calculation made between between 6.2 and 6.6 keV). }
\end{table*}

 \begin{table*}[h]
\caption{\label{tproperties100}
Properties of the sample composed by all the sources with more than 100 counts.
}
 \centering
\begin{tabular}{lrrrrrrl}
\hline\hline
Sample &  $N$  &   $N_{2-12}$  & $N_{5-8}$   & $\langle z \rangle$ & $\langle L_{43} \rangle $  & $\langle N_{h,22} \rangle$    &   Significance \\  
 &   &    &    &  & $ 10^{43} $ erg s$^{-1}$  & $ 10^{22} cm^{-2} $    &   \\ 
 (1)   & (2)  &  (3)    &  (4)    & (5)   &  (6)   &   (7)    &   (8)  \\ 

\hline
 \textbf{Full}     & 219 & 84275  & 19312  & 1.24  & 11.83   & 2.63 & 92 \\
 \hline
\textbf{$\mathrm{log{(N_{\rm H})}}>21.5$}    & 104 & 34893  & 9422  & 1.15  & 10.80   & 5.50  &   91\\
\textbf{$\mathrm{log{(N_{\rm H})}}<21.5$}   & 115 & 49382  & 9890  & 1.32  & 12.69   & 0.04  &  100\\
\hline
\textbf{$\mathrm L_{43} < 6.65 $ }       & 134 &  42061 & 9718  &  0.91 & 2.25   & 3.03   &    91\\

\textbf{ $\mathrm L_{43} > 6.65 $ }     & 85 & 42214  & 9594  & 1.76  & 26.82     & 1.99     &   94  \\
\hline
\textbf{$\mathrm z< 1.01$ }      & 109 & 41477  & 9216  & 0.68  & 3.80   & 2.74 &   90\\

\textbf{$\mathrm z> 1.01$ }      & 110 &  42798 & 10096  & 1.80  & 19.71  & 2.52   &   99\\
\hline
\textbf{$\mathrm  L_{43} < 21.0 $, $\mathrm  z < 0.78$ }  & 71 & 28349  & 6587  & 0.55  & 2.40   & 3.58   &  90 \\

\textbf{$\mathrm  L_{43} < 21.0 $, $\mathrm  z > 0.78$ }       & 93 & 29478  &  6273 & 1.30  &  4.80   & 2.16   &  100\\

\textbf{$\mathrm  L_{43} > 21.0 $, $\mathrm  z > 0.78$ }       &55  & 26448  & 6452  & 2.02  & 35.80  & 2.20   &   90\\

\hline
\end{tabular}
\tablefoot{$L_{43}$: luminosity in units of $10^{43}$ erg s$^{-1}$. Columns: (1): (Sub)Sample; (2): Number of sources; (3): Number of counts in 2 -12 keV; (4): Number of counts in 5-8 keV; (5): average redshift; (6): average luminosity in $10^{43}$ erg s$^{-1}$; (7): average column density of the local absorber in $10^{22}\cdot$ cm$^{-2}$ ; (8): significance of the Fe line calculated as the number of average simulated spectra with a lower flux than the flux of the average observed spectrum (calculation made between between 6.2 and 6.6 keV). }
\end{table*}

\section{Tests of the method: approximation made in the correction of the spectra for detector response}\label{check_unfolding}

We checked whether the correction for the response matrices depends
strongly on the model used to do it.  To make this check, we ran \texttt{XSPEC} simulations.

We start with an ``average'' but high signal-to-noise ratio
  source modelled with an intrinsically absorbed powerlaw and an
  unresolved Gaussian line at 6.4~keV, with the average values of the
  full sample (see Table~\ref{tproperties200}): $z=1.15$,
  $\Gamma=1.24$, $N_{\rm H}=1.75\times10^{22}$~cm$^{-2}$, EW=100~eV.
  We compared the input model with the response-matrix-corrected
  spectra (``unfolded'' spectra in \texttt{XSPEC}) assuming three
  different models:
\begin {enumerate}
\item the input model without the Gaussian
\item a powerlaw with $\Gamma=0$
\item a powerlaw with $\Gamma=2$. 
\end {enumerate}

\begin{figure}
   \centering
   \includegraphics[angle=-90,width=8cm]{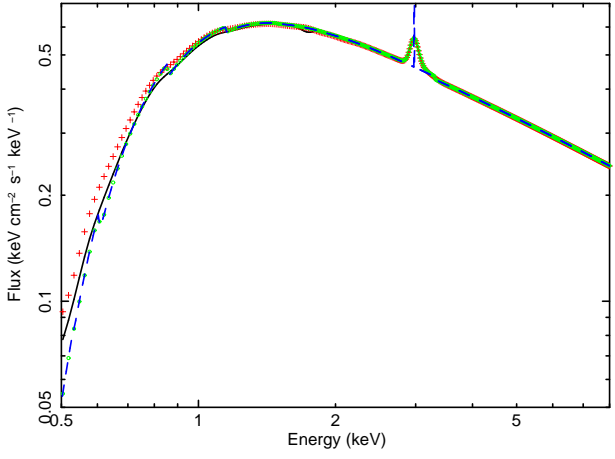}
      \caption{Comparison, for a simulated ``average'' source, between unfolded spectrum with best fit (green circles), with gamma =2 (red plus), with gamma =0 (continuous black line), with model (dashed blue line). 
              }
         \label{Fig:unfoldingcompared}
   \end{figure}

We can see the results in Figs. \ref{Fig:unfoldingcompared} in
  the observed frame.  The best approximation is obtained using the
  input parameters, as expected. However, in a wide range around the
  line, the result does not depend noticeably on the model used for
  the response matrix correction, even for models clearly different
  from the input one. The significant discrepancies start only below
  about 2~keV rest-frame.

\begin{figure}
   \centering
   \includegraphics[angle=-90,width=8cm]{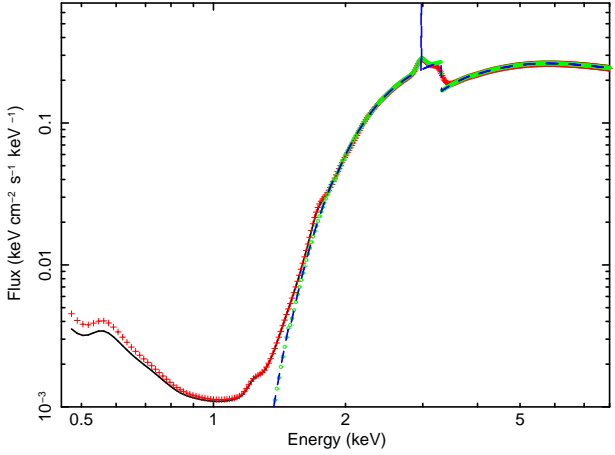}
      \caption{Comparison, for a simulated absorbed source, between: unfolded spectrum with best fit (green circles), with gamma =2 (red plus), with gamma =0 (continuous black line), with model (dashed blue line). The unfolded spectrum obtained with the best fit is along the model, with the exception of the Fe line region, where it deviates.
              }
         \label{Fig:unfoldingcompared_abs}
   \end{figure}

We ran another simulation for an absorbed source, using the
  continuum parameters of one source in the absorbed subsample, ($z=1.15$, $\Gamma=1.64$; $N_{\rm H}=43.22\times10^{22}$~cm$^{-2}$, ) and a Gaussian at 6.4 keV with $\sigma=0$, EW = 100 eV, again with high signal-to-noise ratio. We
  show in Fig.  \ref{Fig:unfoldingcompared_abs} the input model and
  the ``unfolded'' spectra using the three models above. In this case
  we also see that there is no important difference in the continuum
  close to the Fe line between the various models. The differences
  below $\sim$3~keV rest-frame are much stronger in this case compared
  to the ``average'' source, but again the result of the input model
  follows very closely the input spectrum.

Therefore, we chose to correct the spectra for the response
matrices using their own best-fit continuum models and we did
not simply divide the spectra for the effective area of the detector.
This simple method, actually assumes a flat continuum and does not
account for the limited spectral resolution of X-ray detectors.

It is important to notice that in any case there is an energy blending
that widens the Fe line. We have quantified this effect using
  simulations, as explained in Sect. 3.3.

\begin{figure}
   \centering
   \includegraphics[angle=-90,width=8cm]{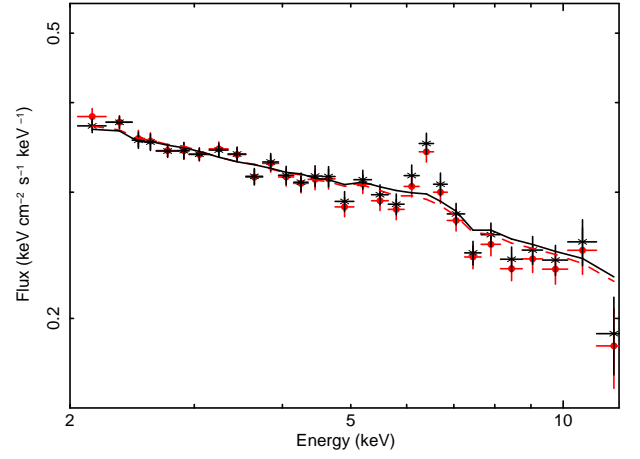}
   \caption{Comparison between the average spectra after having
     corrected for the detector response using: the best fit (average
     observed spectrum represented by full circles, simulated spectrum
     by a dashed line) and with the powerlaw with $\mathrm \Gamma =2 $
     (average observed spectrum represented by stars, simulated
     spectrum by a contiuous line). Sample with the sources with more
     than 200 net counts.  }
         \label{Fig:unfolding_ave200}
   \end{figure}

   In order to check the effect of the correction for detector
   response on the average spectrum, we corrected the spectra for the
   detector responses using a powerlaw with $\Gamma =2 $. We then made
   the average of the observed and simulated spectra, using the same
   procedure explained in the Sect. \ref{method}. We can see in
   Fig. \ref{Fig:unfolding_ave200} the results: the differences
     are again small and, in any case, within the mutual error bars.
     If anything, using the best-fit model for the ``unfolding''
     appears to produce a more conservative estimate of the Fe line
     flux.

\end{document}